\documentclass[reprint,aps,prl,amsmath,amssymb,amsfonts,dvips]{revtex4-1}
\usepackage[dvips]{graphicx,color}
\usepackage{braket,bm}
\usepackage{dcolumn}
\begin{document}

\title{Nature of the randomness-induced quantum spin liquids in two dimensions}

\author{Hikaru Kawamura}
\email[]{kawamura@ess.sci.osaka-u.ac.jp}
\affiliation{Department of Earth and Space Science, Graduate School of Science, Osaka University, Toyonaka, Osaka 560-0043, Japan}

\author{Kazuki Uematsu}
\email[]{uematsu@spin.ess.sci.osaka-u.ac.jp}
\affiliation{Department of Earth and Space Science, Graduate School of Science, Osaka University, , Osaka 560-0043, Japan}

\date{\today}

\begin{abstract}
The nature of the randomness-induced quantum spin liquid state, the random-singlet state, is investigated in two dimensions (2D) by means of the exact-diagonalization and the Hams-de Raedt methods for several frustrated lattices, {\it e.g.\/}, the triangular, the kagome and the $J_1$-$J_2$ square lattices. Properties of the ground state, the low-energy excitations and the finite-temperature thermodynamic quantities are investigated. The ground state and the low-lying excited states consist of nearly isolated singlet-dimers, clusters of resonating singlet-dimers, and orphan spins. Low-energy excitations are either singlet-to-triplet excitations, diffusion of orphan spins accompanied by the recombination of nearby singlet-dimers, creation or destruction of resonating singlet-dimers clusters. The latter two excitations give enhanced dynamical `liquid-like' features to the 2D random-singlet state. Comparison is made with the random-singlet state in a 1D chain without frustration, the similarity and the difference between in 1D and in 2D being highlighted. Frustration in a wide sense, not only the geometrical one but also including the one arising from the competition between distinct types of interactions, play an essential role in stabilizing this {\it frustrated\/} random singlet state. Recent experimental situations on both organic and inorganic materials are reviewed and discussed.
\end{abstract}

%\pacs{}

\maketitle

\section{I. Introduction}

 In the field of magnetism, there has recently been revived interest in the so-called quantum spin liquid (QSL) state \cite{JPSJ-review,book-review,Balents}. The state may vaguely be defined as a possible low-temperature state of strongly interacting magnets, which does not exhibit any kind of magnetic long-range order (LRO) nor a spontaneous symmetry breaking even including the spin-glass-type freezing, as a result of strong quantum fluctuations. The interest was initiated by the proposal of the so-called ``resonating valence bond (RVB)'' state by P.W. Anderson in 1973 \cite{AndersonRVB}. The RVB state was considered as a quantum-mechanical superposition of various types of spin-singlet coverings on a given lattice. Anderson argued that the RVB state might be realized in the $s=1/2$ Heisenberg antiferromagnet on the triangular lattice in which geometrical frustration might help destabilize the Neel-type antiferromagnetic (AF) LRO. Since then, a lot of theoretical and experimental activities were made to realize this hypothetical state in realistic models, and eventually, in real magnets.

 It is relatively recent that promising candidate materials were reported experimentally. In 90's, probably the first candidate of QSL material was reported in the nuclear magnetism of two-dimensional (2D) solid $^3$He film \cite{Fukuyama}. The observed QSL state was gapless accompanied by the temperature($T$)-linear low-$T$ specific heat. In $^3$He, multi-body exchange couplings are generally strong which frustrates the standard bilinear (two-body) exchange coupling.
% between the nearest-neighbor (NN) nuclear spins is ferromagnetic, and the frustration is borne by the competition between the 2-body and the multi-body ($4-, 5-, 6 \dots $-body) couplings. 

 In this century, a variety of QSL candidate materials were reported in more standard electron-spin magnetism, especially in a variety of 2D quantum magnets on geometrically frustrated lattices, {\it e.g.\/}, the triangular and the kagome 
lattices. Well-studied examples are $s=1/2$ organic salts on the triangular lattice-like $\kappa$-(ET)$_2$Cu$_2$(CN)$_3$ \cite{ET-Kanoda,ET-Kawamoto,ET-Kurosaki,ET-Shimizu,ET-Ohira,ET-Nakazawa,ET-Matsuda,ET-Manna,ET-Jawad,ET-Pratt,ET-Goto,ET-Sasaki}, EtMe$_3$Sb[Pd(dmit)$_2$]$_2$ \cite{dmit-Itou,dmit-Tamura,dmit-Itou2,dmit-Matsuda,dmit-Nakazawa,dmit-Itou3,dmit-Watanabe,dmit-Jawad}, $\kappa$-H$_3$(Cat-EDT-TTF)$_2$ \cite{Isono1,Isono2,Ueda}, and $s=1/2$ inorganic kagome antiferromagnet herbertsmithite ZnCu$_3$(OH)$_6$Cl$_2$ \cite{Shores,Helton,Olariu,deVries,Helton2,Freedman,Mendelse,Han,Imai}. These QSL magnets are $s=1/2$ Heisenberg-like magnets with only weak magnetic anisotropy, and exhibit gapless (or nearly gapless) QSL behavior with the $T$-linear low-$T$ specific heat. Similar gapless QSL behavior has also been observed for geometrically-unfrustrated lattices such as the square and the honeycomb lattices where frustration is borne by, {\it e.g.\/}, the competition between the nearest-neighbor (NN) and the next-nearest-neighbor (NNN) interactions, $J_1$ and $J_2$. Examples are a square-lattice magnet Sr$_2$Cu(Te$_{1-x}$W$_{x}$)O$_6$ \cite{Mustonen,Mustonen2,Tanaka} and a honeycomb-lattice magnet 6HB-Ba$_3$NiSb$_2$O$_9$ \cite{Cheng,Darie,Mendels-Ni}. 

 In fact, a similar gapless QSL behavior was reported not only in 2D quantum magnets but also in 3D quantum magnets, {\it e.g.\/}, a highly frustrated pyrochlore magnet, It was recently reported that the mixed-anion $s=1/2$ quantum pyrochlore antiferromagnet Lu$_2$Mo$_2$O$_5$N$_2$ exhibited a gapless QSL behavior characterized by the $T$-linear specific heat \cite{Aattfield}, very much similar to the one observed in various 2D QSL magnets mentioned above.

 Thanks to these active researches, we now have considerable number of experimental realizations of gapless QSL in quantum Heisenberg-like magnets both in 2D and 3D. The true physical origin of such gapless QSL behavior, however, still remains controversial. One promising scenario was proposed by one of the present author (H.K.) and collaborators since 2014. The proposal was that these widely observed gapless QSL-like states might be stabilized by the randomness or the inhomogeneity, either of extrinsic or intrinsic origin \cite{Watanabe,Kawamura,Shimokawa,Uematsu,Uematsu2}. It has been demonstrated in a series of numerical works that such a randomness-induced gapless QSL-like state, called the ``random-singlet state'', is stabilized for 2D random $s=1/2$ Heisenberg models on various frustrated lattices, including the triangular \cite{Watanabe,Shimokawa}, the kagome \cite{Kawamura,Shimokawa}, the $J_1$-$J_2$ honeycomb \cite{Uematsu} and the $J_1$-$J_2$ square \cite{Uematsu2} lattices, as long as the randomness is moderately strong. It has also been argued that the origin of such (effective) randomness in real materials could be of variety, {\it e.g.\/}, the {\it intrinsic\/} ones like the dynamical freezing of the charge (dielectric) degrees of freedom in case of $\kappa$-ET and dmit salts and the slowing down of the proton motion in case of Cat salt, or the {\it extrinsic\/} ones like the possible Jahn-Teller distortion accompanied by the random substitution of Zn$^{2+}$ by Cu$^{2+}$ in case of herbertsmithite and the random occupation of Te/W in case of Sr$_2$Cu(Te$_{1-x}$W$_{x}$)O$_6$.  More recently, the randomness-induced QSL state was confirmed numerically even in 3D for the random $s=1/2$ Heisenberg model on the pyrochlore lattice \cite{Uematsu3}. We note that a pioneering work was also made by Singh, who studied the effect of the site dilution on the ground state of the $s=1/2$ kagome-lattice Heisenberg AF with herbertsmithite in mind, and identified a QSL-like state by means of a series expansion method \cite{Singh}. He called the state the valence-bond glass \cite{Tarzia}.

  In Refs.\cite{Watanabe,Kawamura,Shimokawa,Uematsu,Uematsu2}, the ``random-singlet state'' was vaguely defined as the randomness-induced gapless nonmagnetic state primarily consisting of spin singlets. Its name was invoked from the earlier works on the random AF Heisenberg spin chains in 1D \cite{Ma,Dasgupta,Hirsch,Fisher} and dilute magnetic semiconductors \cite{Bhatt}. In particular, the random-singlet state in 1D was extensively studied by the strong-disorder renormalization-group (RG) method \cite{Ma,Dasgupta,Hirsch,Fisher}. In 1D, the random-singlet state is stabilized without frustration, and is described by an infinite-randomness fixed point \cite{Fisher,Motrunich}. By contrast, the QSL state experimentally observed in the last decade is more tightly-connected 2D and 3D system, and is quite likely to possess {\it a significant amount of frustration\/}, either the geometrical one or the one arising from the competition between distinct types of interactions, {\it e.g.\/}, $J_1$ and $J_2$ \cite{Mustonen,Mustonen2,Tanaka,Cheng,Darie,Mendels-Ni} or the standard two-body exchange and the multi-body exchange \cite{Fukuyama}.

 Hence, the possible similarity and the difference between the random-singlet state studied earlier in the 1D spin chain and the one proposed more recently for frustrated magnets in 2D and 3D remain open. Indeed, the application of the strong-disorder RG to the 2D system did not yield the random-singlet-like fixed point, nor any QSL-like fixed point, but only the spin-glass (SG)-like fixed point \cite{Lin}. Hence, the true relation between the random-singlet state in 1D and the random-singlet state proposed in 2D frustrated magnets (either geometrically frustrated or frustrated via the competition between interactions) remains most interesting, though both are randomness-induced gapless nonmagnetic states.

 Main features of the random-singlet state in 2D and 3D as numerically identified in Refs.\cite{Watanabe,Kawamura,Shimokawa,Uematsu,Uematsu2} are as follows: 1) It is a gapless nonmagnetic state consisting of hierarchically-arranged spin singlets without any characteristic energy-scale other than the interaction energy-scale $J$. 2) The low-$T$ specific heat is proportional to the temperature $T$, apparently due to the nonzero density of low-lying excited states in the low-energy limit $\rho(E\rightarrow 0)> 0$. 3) The susceptibility exhibits a gapless behavior often with a divergent Curie-like tail at lower temperatures, apparently borne by spin-1/2-carrying ``orphan spins'', which are expected to arise reflecting the nontrivial nature of the dimer-covering on the 2D random lattice. 4) The dynamical spin structure factor $S({\bm q}, \omega)$ exhibits broad features both in $q$ and in $\omega$, without any gap nor distinct peak structure. All these features are observed more or less in common in all the 2D and 3D models studied in Refs.\cite{Watanabe,Kawamura,Shimokawa,Uematsu,Uematsu2,Uematsu3}, and seem to be a universal character of the randomness-induced QSL state. The state might be described as an ``Anderson-localized resonating-valence bond (RVB) state''. Or, if one considers the fact that the singlet-dimer localization here is essentially a many-body effect, one might better say a ``many-body localized RVB state''. Anyway, the global resonance in the RVB state is changed into the local resonance in the random-singlet state.

 It should also be noticed that the random-singlet state in Refs.\cite{Watanabe,Kawamura,Shimokawa,Uematsu,Uematsu2,Uematsu3} was best described for the case of strong randomness, simply because the properties of the randomness-induced QSL state, the random-singlet state, tend to be most eminent in the strong random case, clearly manifesting themselves even in smaller systems accessible by the ED method employed in Refs.\cite{Watanabe,Kawamura,Shimokawa,Uematsu,Uematsu2,Uematsu3}. How the random-singlet state extends to weaker randomness, and how it looks like there, remains an interesting question. 

 More recently, other groups also studied the effect of randomness (disorder) on the possible QSL behavior in certain 2D quantum spin models \cite{Kimchi,Kimchi2,Guo,Sheng}. For example, Wu, Gong and Sheng performed a density-matrix renormalization-group (DMRG) analysis of the $s=1/2$ $J_1$-$J_2$ Heisenberg AF on the triangular lattice, extending the earlier ED calculation of Ref.\cite{Watanabe,Shimokawa} to include the NNN interaction $J_2$, and to slightly larger lattices of $N\leq 48$ ($N\leq 30$ in the ED calculation of Refs.\cite{Watanabe,Shimokawa}), and confirmed the appearance of the randomness-induced spin-liquid-like state, without any indication of the SG state \cite{Sheng}.

The weaker random case, which might be realized, {\it e.g.\/}, close to the valence-bond-crystal (VBC) order, was analyzed in Refs.\cite{Kimchi} and \cite{Guo}. Liu, Shao, Lin, Guo and Sandvik performed a quantum Monte Carlo (QMC) analysis of a special type of $s=1/2$ Heisenberg AF on the square lattice in which a six-body interaction frustrates the standard NN AF bilinear interaction. These authors succeeded in avoiding the notorious negative sign problem inherent to most of frustrated systems, and could treat much larger systems than those  studied by the ED and DMRG methods. Although the model was classified as an unfrustrated model in Ref. \cite{Guo}, it might be better classified as a frustrated model from a physical point of view, since the six-body interaction frustrates the NN bilinear coupling in the sense that the two interactions favor distinct types of orderings, exactly the same situation occurring in, {\it e.g.\/}, the $J_1$ and $J_2$ interactions in the square-lattice model of Ref.\cite{Uematsu2}.
Kimchi, Nahum and Senthil, and also Liu {\it et al\/}, emphasized the importance of ``the defect carrying the spin-1/2'' \cite{Kimchi}, or ``the spinon'' \cite{Guo} appearing at the nexus of different VBC-order domains, and regarded it as essential ingredients of the random-singlet order. Although Liu {\it et al\/} mentioned that such a spinon excitation was neglected in the ED analysis of Refs.\cite{Watanabe,Kawamura,Shimokawa,Uematsu,Uematsu2}, it seems to the present authors that such a ``spinon'' or a ``spin-1/2 carrying defect'' might in fact be closely related to the ``orphan spin'' discussed  as an essential ingredient of the random-singlet state in Refs.\cite{Watanabe,Kawamura,Shimokawa,Uematsu,Uematsu2}. Presumably, the former object best described in the weak random case might take the form of the latter in the strong random case, though the precise correspondence needs further clarification. 

 In this way, there is now a considerable amount of both experimental and theoretical developments achieved in the randomness-induced QSL behavior in 2D and 3D quantum magnets, yet still many things need to be clarified and understood. Under such circumstances, the purpose of the present paper is to further clarify the nature of the 2D random-singlet state proposed in Refs.\cite{Watanabe,Kawamura,Shimokawa,Uematsu,Uematsu2} by examining the microscopic character of the ground state and the low-lying excited states by means of the exact-diagonalization (ED) method. We study several 2D frustrated spin models sustaining the random-singlet state, including the triangular, the kagome and the $J_1$-$J_2$ square lattice models. Comparison is made with the unfrustrated 1D random Heisenberg chain  to get deeper insights into the random-singlet state in 2D frustrated magnets by clarifying the similarity to and the difference from the random-singlet state in 1D. 

 The subsequent part of the paper is organized as follows. In \S II, we define the model and explain the numerical method employed. The nature of the 2D random-singlet ground state examined by means of the ED method is presented and is compared with that of the 1D model in \S III. The nature of the low-energy excitations is examined in comparison with that of the 1D model in \S IV. Thermodynamic properties such as the specific heat and the susceptibility of the 2D random-singlet state at finite temperatures are examined in comparison with those of the 1D model in \S V. Finally, \S VI is devoted to summary and discussion.

\section{II. The model}

The models we study are basically the same models as studied earlier in Refs.\cite{Watanabe,Kawamura,Shimokawa,Uematsu,Uematsu2}, {\it i.e.\/}, the random-bond $s=1/2$ isotropic Heisenberg model on several frustrated 2D lattices, including the triangular, the kagome and the $J_1$-$J_2$ square lattices. The Hamiltonian is given by
\begin{align}
\mathcal{H}=J_1\sum_{\Braket{i,j}}j_{ij}{\bm S}_i\cdot{\bm S}_j +
J_2\sum_{\Braket{\Braket{i,j}}} j_{ij}{\bm S}_i\cdot{\bm S}_j,
\label{eq:hamiltonian}
\end{align}
where ${\bm S}_i=(S_i^x, S_i^y, S_i^z)$ is the $s=1/2$ spin operator at the $i$-th site on the lattice, the sums $\Braket{i,j}$ and $\Braket{\Braket{i,j}}$ are taken over all NN and NNN pairs on the lattice, where both the NN coupling $J_1$ and the NNN coupling $J_2$ are assumed to be AF, {\it i.e.\/}, $J_1>0$ and $J_2>0$, and we put $J_1=1$ as an energy unit. Randomness of the model is represented by the variable $j_{ij}$ which is the random variable obeying the bond-independent uniform distribution between $[1-\Delta, 1+\Delta]$ with $0\leq \Delta\leq 1$. The parameter $\Delta$ represents the extent of the randomness: $\Delta=0$ corresponds to the regular case and $\Delta=1$ to the maximally random case so long as the interaction is restricted to be AF.  The extent of the randomness $\Delta$ is taken to be common between $J_1$ and $J_2$. In the cases of triangular and the kagome lattices where geometrical frustration arises even for the $J_1$-only model, we put $J_2=0$, while for the case of the square lattice, we put a nonzero $J_2>0$ to introduce frustration into the model. Periodic boundary conditions are applied in both directions.

 For comparison, we also study the 1D random-bond $s=1/2$ Heisenberg model described by the same Hamiltonian (\ref{eq:hamiltonian}). The random-singlet state in 1D has been discussed in literature mostly for the case of $J_1$-only model where the frustration is totally absent \cite{Ma,Dasgupta,Hirsch,Fisher}. In order to highlight the possible difference from and the similarity to this well-studied case, we put in the present paper $J_1>0$ and $J_2=0$ in the 1D model, following the earlier works.

 The properties of the ground-state and the excited state of the model are computed by the ED Lanczos method ($T=0$) and by the Hams-de Raedt method ($T>0$) \cite{HamsRaedt}. The Hams-de Raedt method obtains the finite-temperature properties by replacing the thermal average with the average over a few thermally typical pure states produced via the imaginary time evolution of initial random vectors. We treat finite-size clusters with the total number of spins $N$ up to $N\leq32$ both in the Lanczos and the Hams-de Raedt methods. The total number of samples are $N_s=100$ for $N\leq 24$, $N_s=30$ for $N=30$, and $N_s=10$ for $N=32$ for all the lattices studied including the 1D one.

\section{III. The nature of the ground state}

In this section, we examine the microscopic character of the ground-state wavefunctions of the $s=1/2$ random-bond Heisenberg model on several frustrated 2D lattices, including the triangular, the kagome and the $J_1$-$J_2$ square lattices. In the case of the $J_1$-$J_2$ square model, we put $J_2=0.5$ where the random-singlet state is stabilized in a wide parameter region \cite{Uematsu2}. The random-singlet state turned out to be the ground state of the model when the extent of the randomness $\Delta$ exceeds a critical value $\Delta_c$, $\Delta_c$ being estimated to be $\sim 0.5$ (triangular), $\sim 0.4$ for (kagome) and $\sim 0.6$ ($J_1$-$J_2$ square). In the following, in order to see the typical behavior of the random-singlet ground state, we mainly consider the case of the maximal randomness of $\Delta=1$ where the properties of the random-singlet state is expected to be most eminent even for small sizes accessible by the present ED method.

 The random-singlet state is basically a state consisting of hierarchically-arranged spin-singlet dimers. An indicator of the `strength' of the singlet might be the ground-state expectation value $\langle {\bf S}_i\cdot {\bf S}_j\rangle$. For an isolated spin pair, $\langle {\bf S}_i\cdot {\bf S}_j\rangle=-3/4$ for the singlet state with $S=0$ ($S$ the total spin), while $\langle {\bf S}_i\cdot {\bf S}_j\rangle=1/4$ for the triplet state with $S=1$, and $\langle {\bf S}_i\cdot {\bf S}_j\rangle$ generally takes a value between [$-3/4,\ 1/4$]. 

\begin{figure}
  \begin{center}
    \includegraphics[width=\hsize]{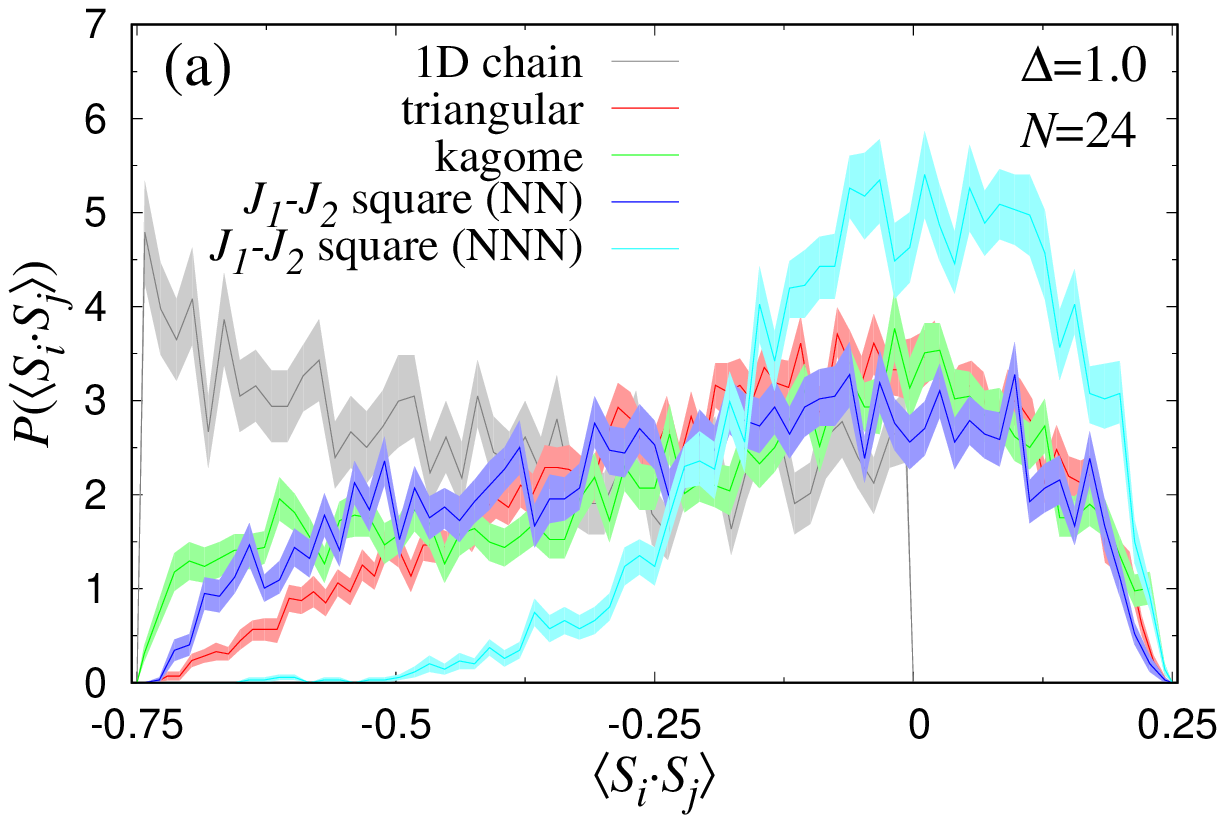}
    \includegraphics[width=\hsize]{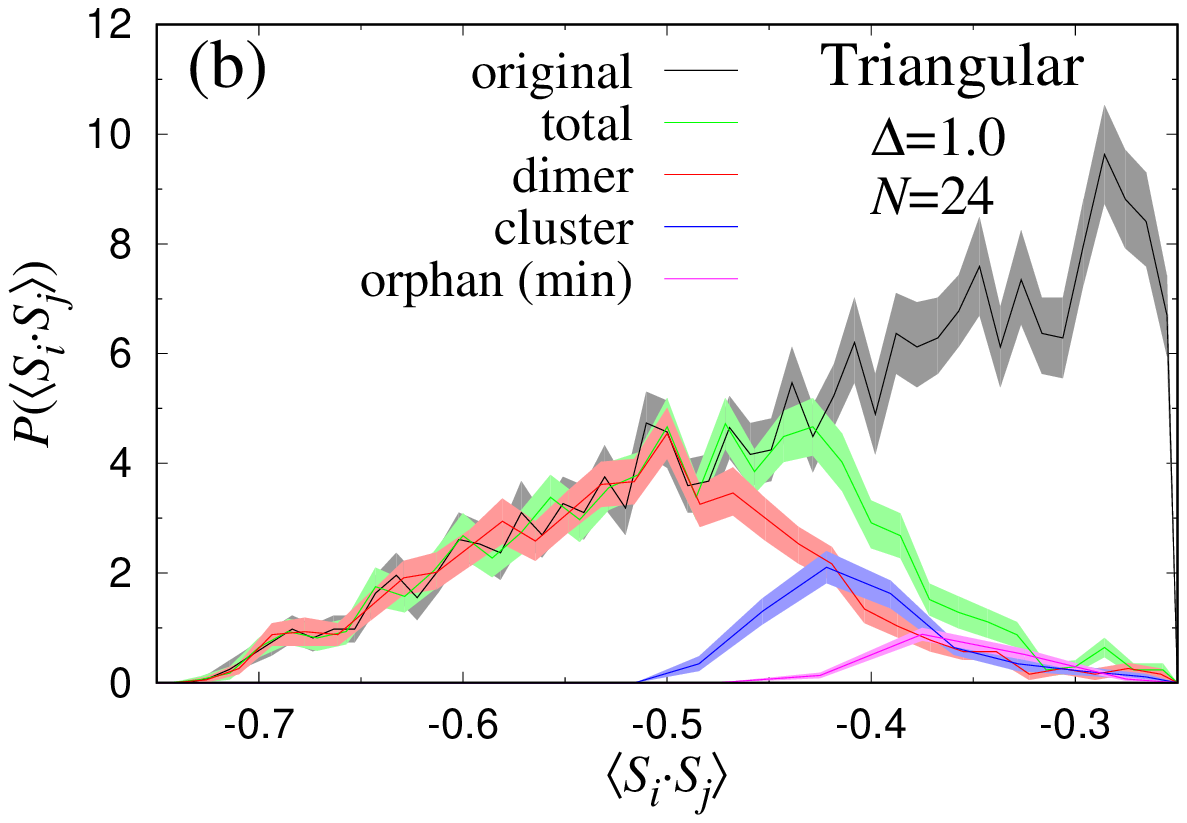}
   \caption{\label{fig:DSF} 
(Color online) (a) The distribution of the ground-state expectation value $\langle {\bf S}_i\cdot {\bf S}_j\rangle$ of all NN bonds on the triangular, the kagome, the $J_1$-$J_2$ ($J_2=0.5$) square, and the 1D lattices. In case of the $J_1$-$J_2$ square lattice, the plots are given also for the NNN bonds on which the interaction $J_2=0.5$ works. (b) The distribution of the ground-state expectation value $\langle {\bf S}_i\cdot {\bf S}_j\rangle$ for arbitrary $<ij>$ pairs in the case of the triangular lattice, obtained after the singlet-dimer formation procedure explained in the text, each for isolated singlet-dimers, singlet-dimers clusters, and orphan spins, in comparison with the original distribution. For orphan spins, the distribution is shown only for the minimum $\langle {\bf S}_i\cdot {\bf S}_j\rangle$ for given orphan spin. In both (a) and (b), the lattice size is $N=24$.
} 
  \end{center}
\end{figure}

 In Fig.1, we show the distribution of $\langle {\bf S}_i\cdot {\bf S}_j\rangle$ for all NN bonds on the lattice and for all $N_s$ samples studied, for the case of the triangular, the kagome, the $J_1$-$J_2$ square lattices. For the the $J_1$-$J_2$ ($J_2=0.5$) square lattice, the $\langle {\bf S}_i\cdot {\bf S}_j\rangle$ distribution on the NNN bonds is also given. For comparison, the corresponding distribution of the 1D chain is given. The distribution in the 2D cases looks more or less similar to each other, whereas the distribution in the 1D case exhibits a relatively large weight on the smaller $\langle {\bf S}_i\cdot {\bf S}_j\rangle$-value, probably because the present 1D model has no frustration.
 
 The optimal singlet-dimer configuration is achieved to minimize the total energy for a given random distribution of the exchange couplings $\{J_{ij}\}$. Roughly speaking, strong singlets would be formed on strong $J_{ij}$-bonds, while weak singlets on weak $J_{ij}$-bonds. However, such a dimer-covering problem on the random lattice is highly nontrivial even for the NN model. 

 \begin{figure}[t]
  \includegraphics[width=7cm]{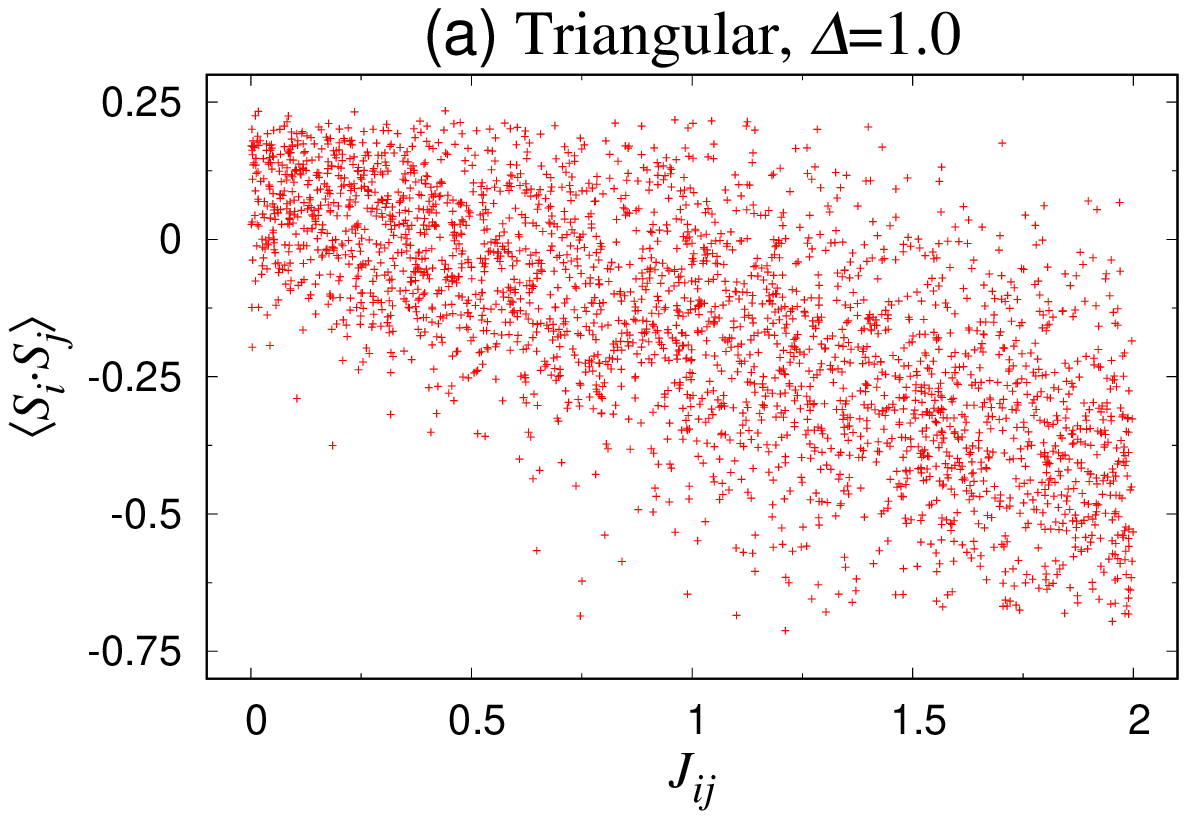}\\
  \includegraphics[width=7cm]{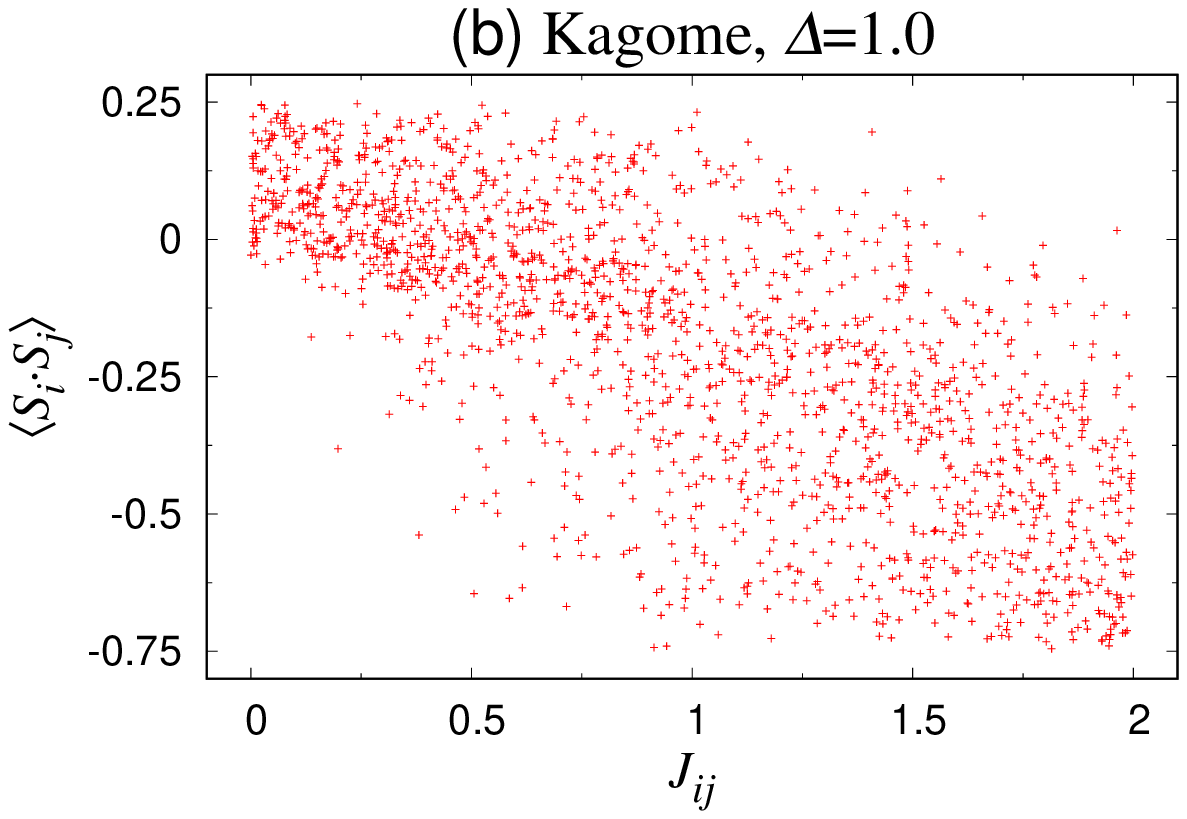}\\
  \includegraphics[width=7cm]{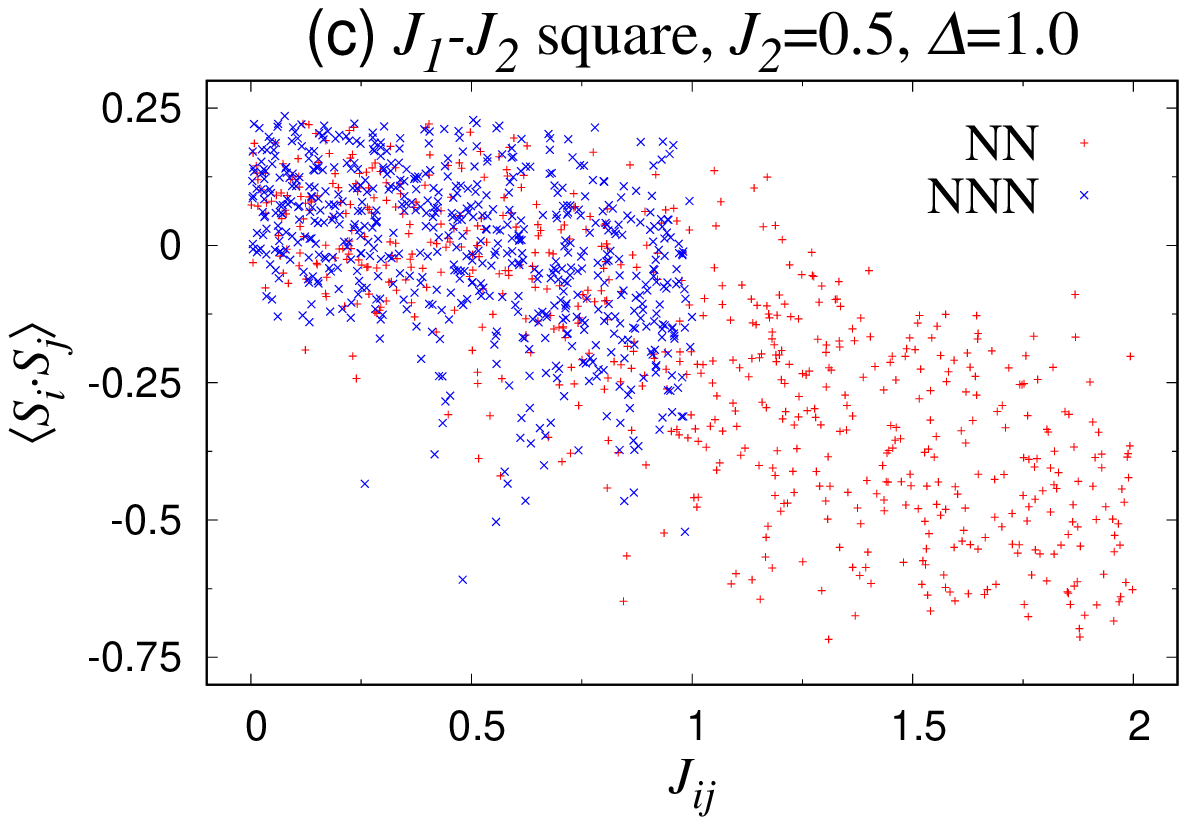}\\
  \caption{\label{fig:order} 
(Color online) The ground-state expectation value $\langle {\bf S}_i\cdot {\bf S}_j\rangle$ plotted versus $J_{ij}$ for all NN bonds on (a) the triangular, (b) the kagome, and (c) the $J_1$-$J_2$ ($J_2=0.5$) square lattices of the sizes $N=30$ (triangular and kagome) and $N=32$ ($J_1$-$J_2$ square). In case of the $J_1$-$J_2$ square lattice, the plots are given also for the NNN bonds at which the interaction $J_2$ works. 
} 
\end{figure}

 Thus, in Fig.2, we plot $\langle {\bf S}_i\cdot {\bf S}_j\rangle$ versus $J_{ij}$ for all bonds and for all samples, for the case of (a) the triangular, (b) the kagome and (c) the $J_1$-$J_2$ ($J_2=0.5$) square lattices. Clearly, there is a correlation between the $J_{ij}$-value and the $\langle {\bf S}_i\cdot {\bf S}_j\rangle$-value, {\it i.e.\/}, a tendency of strong singlets formed on strong bonds and weak singlets on weak bonds. Nevertheless, there is a considerable deviation from such a tendency, especially in the physically interesting region of larger $J_{ij}$. More quantitatively, the correlation coefficient is estimated in the stronger $J_{ij}$-regime of $J_{ij}\geq 1$ to be 0.5 for (a), 0.4 for (b), and 0.2 for (c). This demonstrates the nontrivial nature of the singlet-dimer formation in the random-singlet state.

 In order to get more detailed microscopic information on each sample, we try to construct from the numerically determined exact ground state for a given $J_{ij}$ distribution a singlet-dimer configuration according to the following procedure. Let $e_{ij}$ be the value of $\langle {\bf S}_i\cdot {\bf S}_j\rangle$ for an arbitrary spin pair ($ij$). First, all possible bonds (spin pairs) ($ij$) including all distant-neighbor bonds satisfying the threshold condition $e_{ij} < e_c$, are ordered according to their $e_{ij}$-values from smaller ones (negative values with their absolute values large) to larger ones, where $e_c$ is a threshold value. According to this order, we successively regard the bonds as forming ``singlet-dimers'', with imposing a constraint that a site $i$ which has been involved in already assigned singlet-dimers can no longer be allowed to be included in a new singlet-dimer, {\it i.e.\/}, any site cannot be involved simultaneously in more than one singlet-dimers. The picture behind such a singlet-dimer formation procedure is that, under the random-$J_{ij}$ environment, a given site tends to form a more or less tightly-bound singlet-dimer with a particular other site, {\it i.e.\/}, the random-singlet picture of the 1D model \cite{Fisher}. In the following calculation, we take the threshold value $e_c=-0.25$. The reason behind this choice is that, for the two spins with $e_{ij}\geq -0.25$, their ``entanglement of formation'' (or ``concurrence'') vanishes in the present $SU(2)$ limit, meaning the two spins are disentangled \cite{Bennet,Hill,Wootters,OConnor}. 

 Meanwhile, reflecting the intrinsic frustration effect associated with the dimer-covering problem on the random lattice, there could occur a ``resonance'' among distinct dimer-coverings. In such a case, a particular site $i$ might simultaneously be involved into two different bonds ($ij$) and ($ik$) with nearly common values of $e_{ij}$ and $e_{ik}$ via such a resonance process. Such a local resonance of singlet-dimers tends to lead to a ``singlet-dimers cluster'' consisting of more-than-two spins. In order to take account of such a ``resonance'' effect, we introduce a modification of the above rule as follows.

When the sites $i$ and $j$ form a new singlet-dimer with a certain $e_{ij}<e_c$ value according to the above rule, we examine the $e$-values associated with the bond between the site $i$ (or $j$) and an other site $k(\neq i,j)$ not yet involved in any pre-formed singlet-dimer. If this $e_{ik}$ (or $e_{jk}$)-value happens to be close to the $e_{ij}$-value, satisfying the condition $e_{ik}-e_{ij}<\delta$ (or $e_{jk'}-e_{ij}<\delta$), the ($ik$) (or ($jk'$)) bond is regarded as a singlet-dimer, which leads to the formation of a more-than-2-spin cluster. This procedure is repeated for the newly added site $k$ (or $k'$). The threshold value $\delta$ is somewhat arbitrary, but we take it $\delta=1/32$ in the following procedure. 

 Such a clustering process can further be generalized as follows. Suppose that, during the clustering procedure, we have a $m$-spin cluster with its maximum (absolute-value minimum) $e_{ij}$-value ($i,j\in$ the $m$-spin cluster) $e_{max}$. Let $i_c$ be an arbitrary site belonging to this $m$-spin cluster, and $k$ a site which is either an arbitrary site belonging to this cluster ($k\neq i_c$) or an unpaired site outside the $m$-spin cluster. We examine for all ($i_ck$) bonds whether the condition $e_{ik}-e_{max}<\delta$ is satisfied or not, and if it is satisfied, we regard the ($i_ck$) bond as a new singlet-dimer and include it into the $m$-spin cluster. As long as any new singlet-dimer is added to the cluster, we repeat the process with a new $e_{max}$. When there is no new singlet-dimer added to the cluster, the clustering process on this $m$-spin cluster is finished, and we return to the list of ordered $\{e_{ij}\}$ at the maximum $e_{ij}$-bond left thus far, and continue the singlet-dimer formation procedure until our list of $\{ e_{ij}(<e_c) \}$ is exhausted.

 When this procedure is complete, some fraction of sites are still left unpaired with any other site, not involved in any singlet-dimer. We call these ``left-over'' spins failing to form singlet-dimers ``orphan spins''. Since we have taken $e_c=-0.25$, ``orphan spins'' by this definition are disentangled with all other spins.

 In Fig.1(b), we show the distribution of $e_{ij}$ for arbitrary $<ij>$ pairs obtained after the above-mentioned singlet-dimer formation procedure in the case of the triangular model, each for isolated singlet-dimers, singlet-dimers clusters, and orphan spins, together with the original $e_{ij}$-distribution. For orphan spins, the distribution of the minimum $e_{ij}$ for any given orphan spin is shown. As can be seen from Fig.1(b), the distribution at smaller $e_{ij}$ is dominated by isolated singlet-dimers. For somewhat larger $e_{ij}$, resonating singlet-dimers clusters begin to have some weights, while orphan spins have appreciable weights only for still larger $e_{ij}$.

 \begin{figure}[t]
  \includegraphics[width=6.5cm]{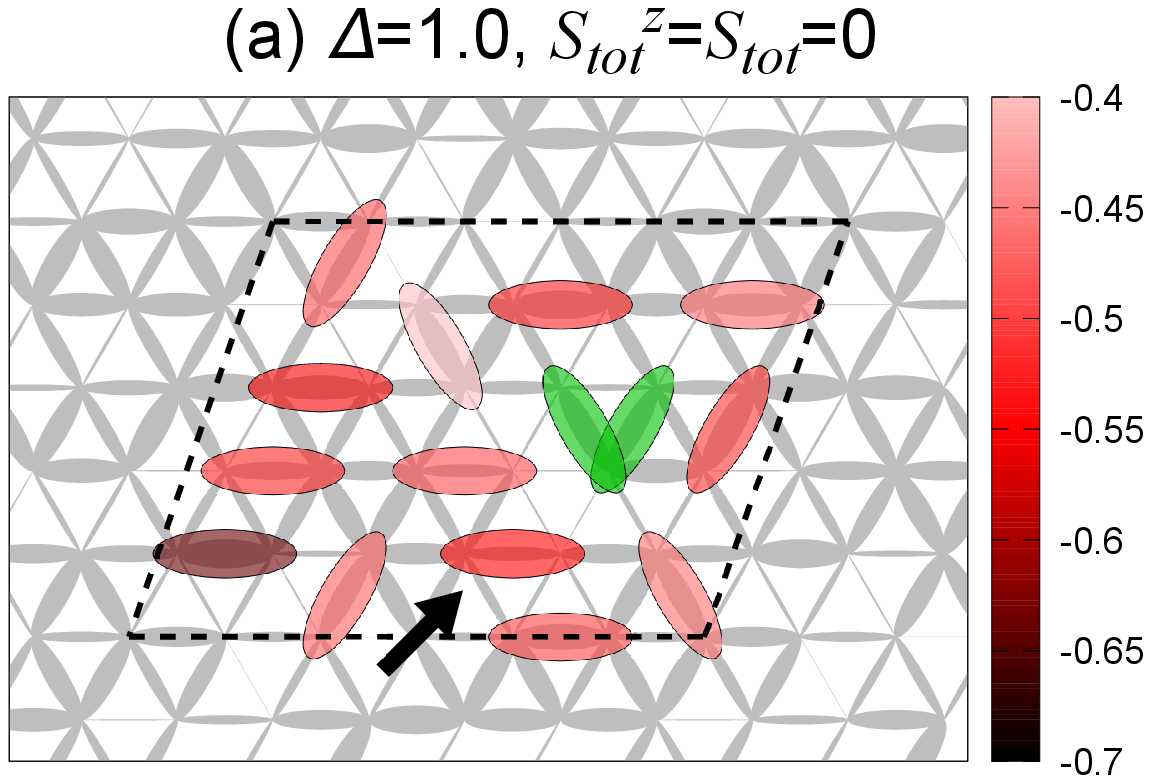}\\
  \includegraphics[width=5.5cm]{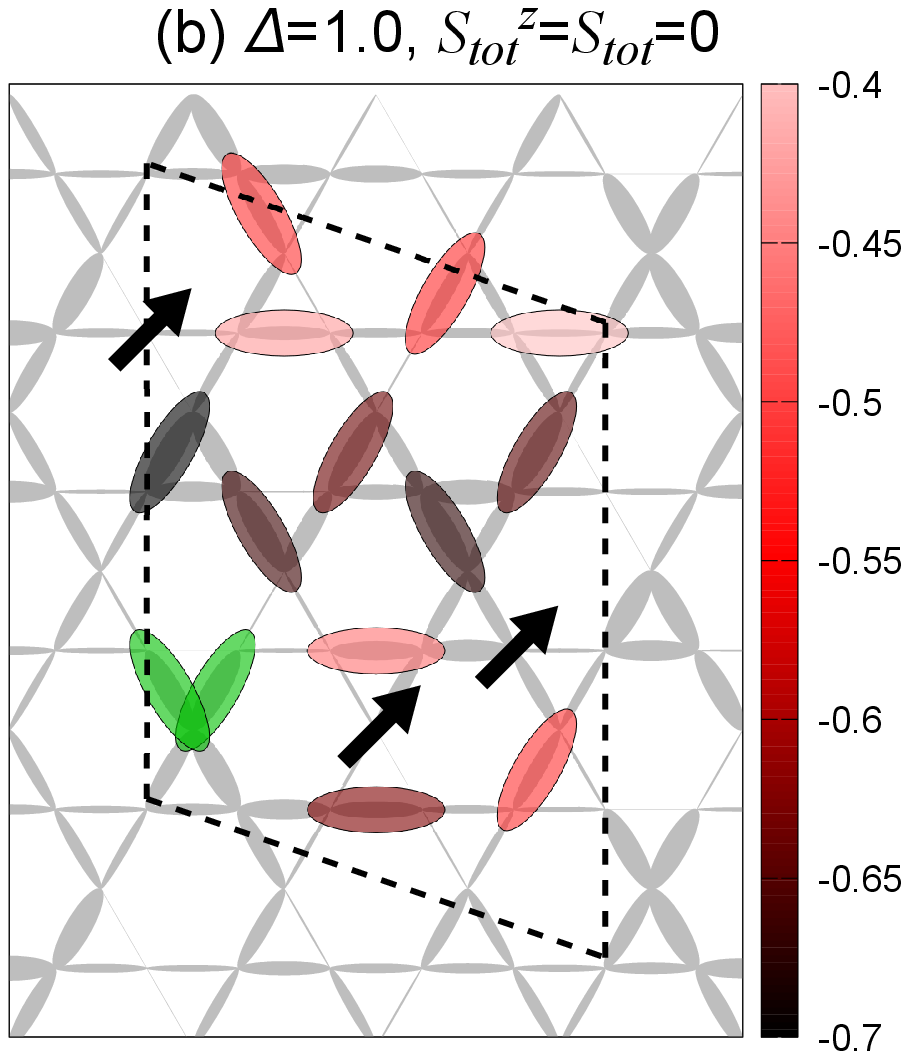}\\
  \includegraphics[width=7cm]{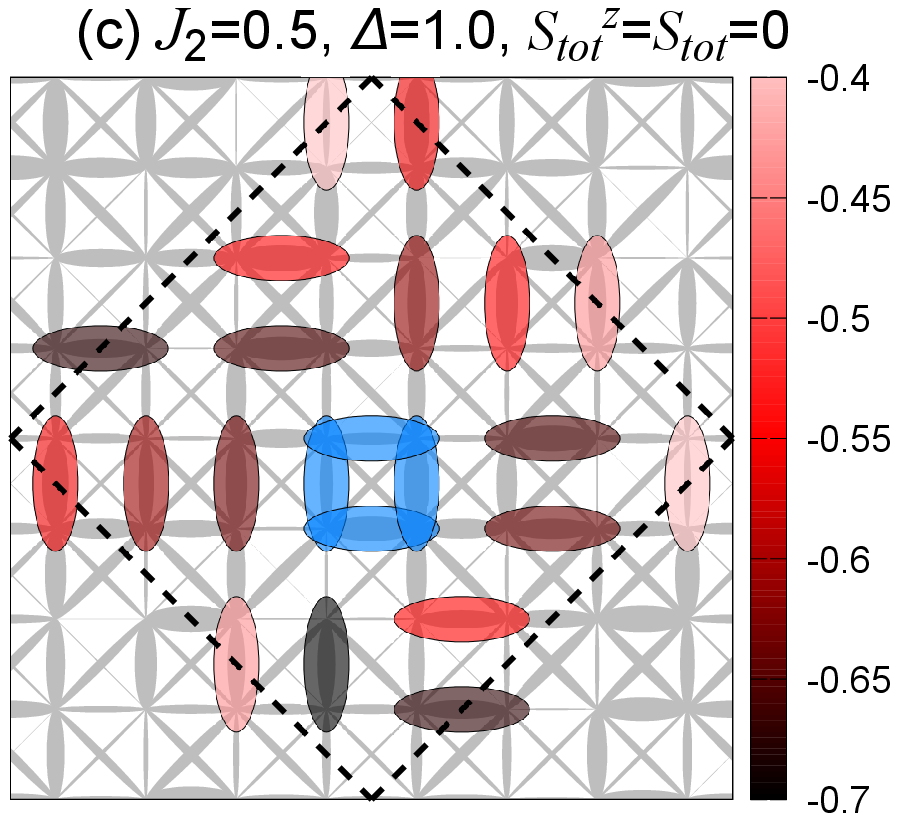}\\
  \caption{\label{fig:order} (Color online) 
Typical singlet-dimer configurations of the singlet ($S=0$) ground state of (a) the triangular, (b) the kagome, and (c) the $J_1$-$J_2$ ($J_2=0.5$) square lattices of the size $N=30$ (triangular and kagome) and $N=32$ ($J_1$-$J_2$ square). Ellipses and arrows represent singlet-dimers and orphan spins, respectively. Gray bonds represent the interaction $J_{ij}$, whose thickness represents its strength.
} 
\end{figure}

As an example, we show in Fig.3 typical singlet-dimer configurations obtained in the way explained above for exact ground states of (a) the triangular, (b) the kagome and (c) the $J_1$-$J_2$ ($J_2=0.5$) square lattices, for their typical singlet ($S=0$) ground states. Red ellipses in these figures represent isolated singlet-dimers, its brightness representing the associated $e_{ij}$-value, {\it i.e.\/}, the $\langle {\bf S}_i\cdot {\bf S}_j\rangle$-value. Most of these red singlet-dimers are formed between NN spins, but some are between further-neighbor spins. Blue (green) clusters consisting of more than one singlet-dimers represent the resonating singlet-dimers cluster described above, a quantum-mechanical superposition of more than one singlet-dimers (and orphan spin) configurations. Arrows in the figure represent orphan spins explained above. Thus, all sites (spins) on the lattice should belong to either (i) isolated singlet-dimers, (ii) resonating singlet-dimers clusters, or (iii) orphan spins.

  In the range of sizes studied, the random-singlet ground states of the 2D random Heisenberg model are mostly singlets ($S=0$) and some fraction of triplets ($S=1$), with a very small fraction of quintets ($S=2$) also observed in the kagome and the triangular cases. While the singlet-dimer configurations shown in Fig.3 are the ones of the singlet ground states, we find no appreciable difference between the singlet-dimer configurations of the singlet and of the triplet (and even the quintet) ground states.

 In Fig.4, similar singlet-dimer configurations are shown also for typical triplet ($S=1$) ground states in each case of (a) the triangular, (b) the kagome and (c) the $J_1$-$J_2$ ($J_2=0.5$) square lattices. In the figures, we also include the spatial spin-density $\langle S_z\rangle $-distribution of the corresponding $S_z=1$ sample. Note that, since our model is a zero-field model with the time-reversal symmetry, any state with $S_z=0$, including all singlet states shown in Fig.3, gives a trivial distribution with $S_{iz}=0$ for all $i$. By contrast, triplet states with $S_z=\pm 1$ yield nontrivial $\langle S_z \rangle$-distribution as shown in Fig.4. Bias toward the positive $\langle S_z\rangle$-value in Fig.4 is simply due to our choice of the $S_z=1$ ground state: Opposite negative bias appears if one takes the $S_z=-1$ ground state. 

 Some information about the ground state can be obtained from the local $\langle S_z\rangle$-distribution shown in Fig.4 for the triplet ($S=1,S_z=1$) sample. The $\langle S_z\rangle$-value tends to be large for orphan spins and small for isolated singlet-dimer forming spins, though there are some exceptions to this tendency. When appreciable spin density appears at an isolated singlet-dimer, the associated $e_{ij}$-value tends to be not so small, and the induced local spins on the two sites belonging to this dimer tend to point opposite exhibiting an AF correlation. For the singlet-dimer with its $e_{ij}$-value very small (or $|e_{ij}|$-value very large), by contrast, nearly-vanishing spin density is generated.

 \begin{figure}[t]
  \includegraphics[width=6.5cm]{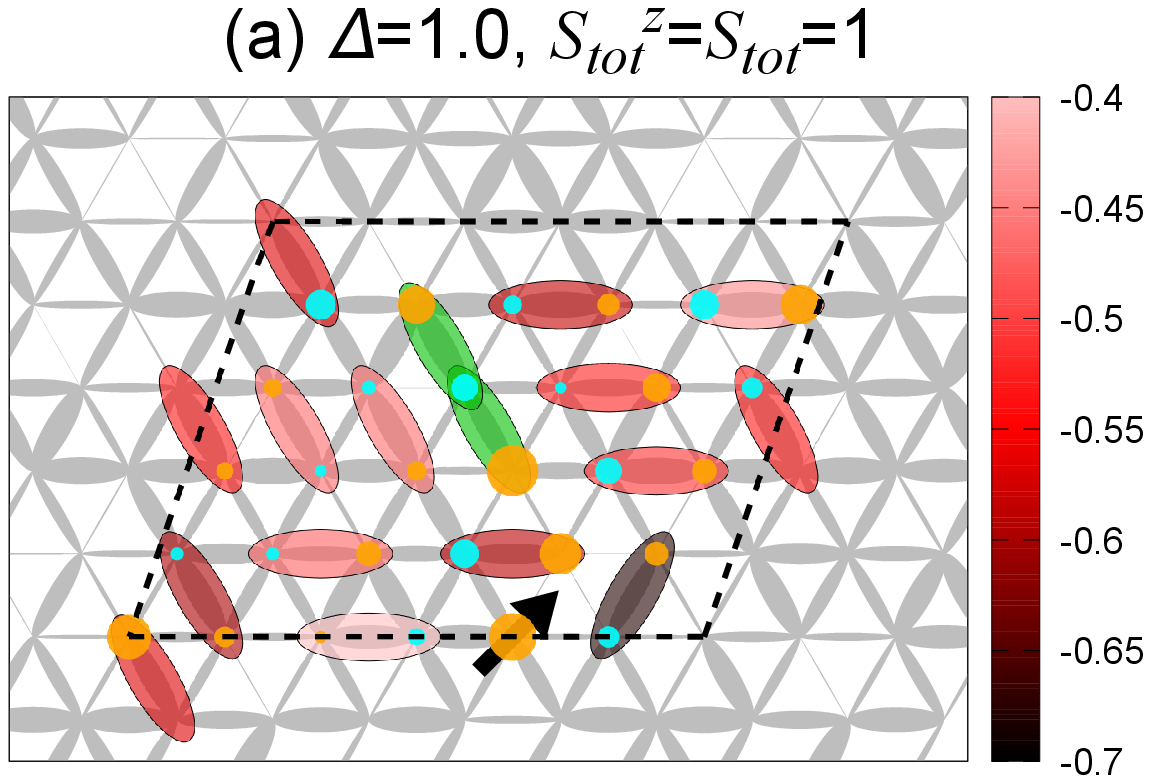}\\
  \includegraphics[width=5.5cm]{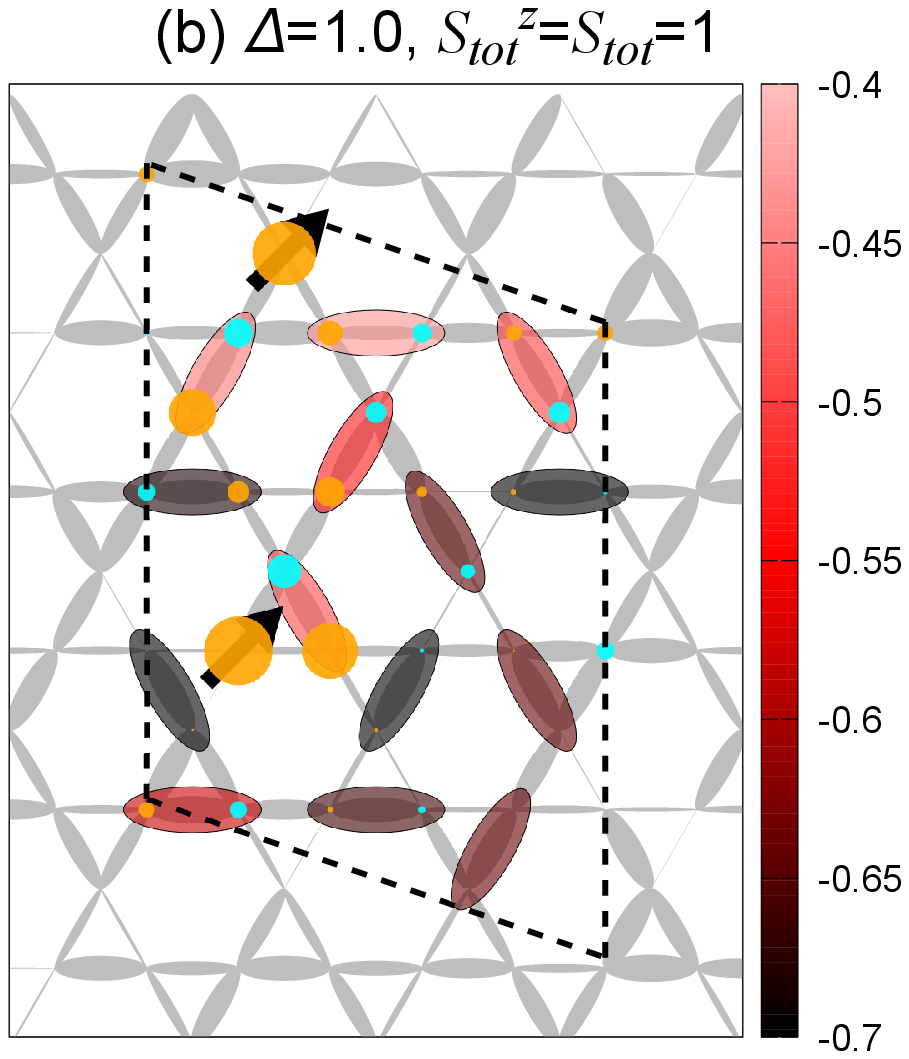}\\
  \includegraphics[width=7cm]{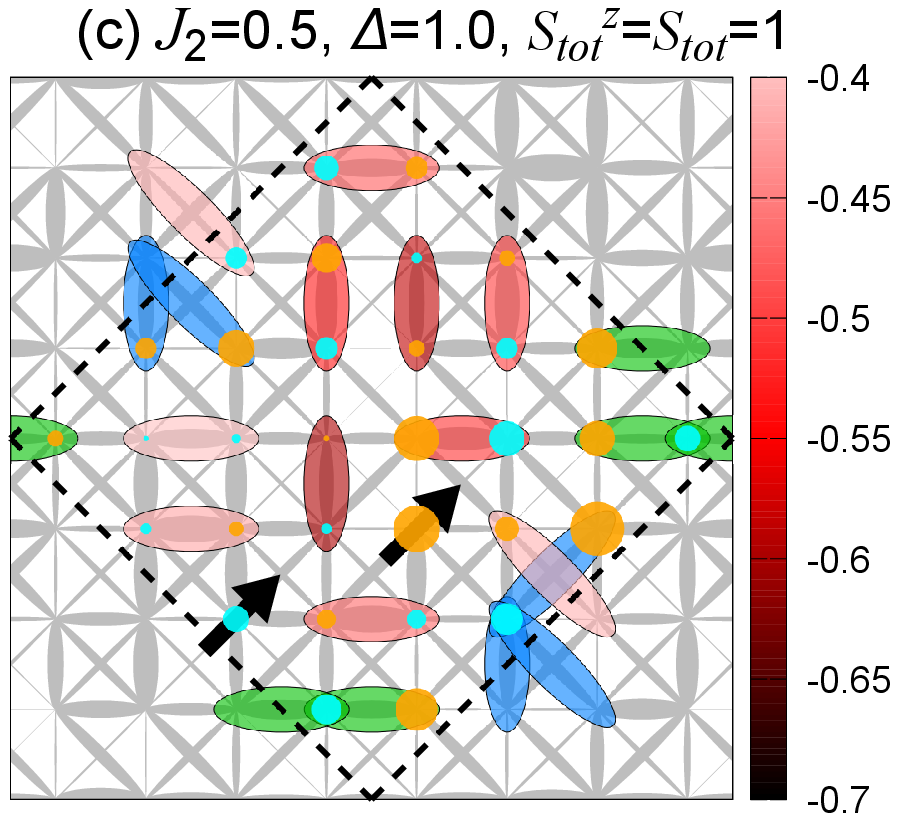}\\
  \caption{\label{fig:order} (Color online) 
Typical singlet-dimer configurations of the triplet ($S=S_z=1$) ground state of (a) the triangular, (b) the kagome, and (c) the $J_1$-$J_2$ ($J_2=0.5$) square lattices of the sizes $N=30$ (triangular and kagome) and $N=32$ ($J_1$-$J_2$ square). Ellipses and arrows represent singlet-dimers and orphan spins, respectively, while yellow (blue) circles represent the local spin density $\langle S_z\rangle$ of up (down) spin, its size representing the magnitude. Gray bonds represent the interaction $J_{ij}$, whose thickness represents its strength.
} 
\end{figure}

 In Fig.5, we show the ratio of each type of configurations (i)-(iii) above, {\it i.e.\/}, (i) isolated singlet-dimers, (ii) resonating singlet-dimers clusters, and (iii) orphan spins, averaged over all $N_s$ samples including both singlet and triplet ground states, for the cases of the triangular, the kagome and the $J_1$-$J_2$ ($J_2=0.5$) square lattices. The corresponding result for the 1D NN chain is also given. The system-size dependence is shown by comparing the data for $N=16$ (or 18), 24, and 30 (or 32). As can clearly be seen from the figure, the size dependence seems to be rather minor. Furthermore, the ratio pattern looks more or less similar among different 2D lattices. The random-singlet state in 2D consists primarily of isolated singlet-dimers, with some fraction of resonating singlet-dimers clusters and of orphan spins, each of 10\% and 5\% order. More precisely, fractions of isolated singlet-dimers, resonating singlet-dimers clusters and orphan spins are (79\%, 14\%, 6\%), (84\%, 9\%, 7\%), and (89\%, 9\%, 3\%) for the triangular, the kagome and the $J_1$-$J_2$ square lattices, respectively,  for the largest size studied, {\it i.e.\/}, $N=30$ for the triangular and the kagome lattices and $N=32$ for the $J_1$-$J_2$ square lattice.

 In 1D, by contrast, the fractions of resonating singlet-dimers clusters and of orphan spins are reduced somewhat while the fraction of isolated singlet-dimers are enhanced, the corresponding fractions being (92\%, 5\%, 3\%) for $N=32$. While the isolated singlet-dimers in 2D are formed mostly between the NN sites, with the exception of the NNN sites of the $J_1$-$J_2$ square lattice at which the interaction works directly, the rate of further-neighbor isolated singlet-dimers is relatively increased in 1D, around 9\%, significantly greater than the corresponding rate of the 2D lattices of order 1\% or even less. Hence,  there exists some appreciable difference between the appearance patterns of the singlet-dimer configurations in 2D and in 1D. %This probably reflects the fact that the frustration and the associated degeneracy of states are more tense in 2D than in 1D.

 \begin{figure}[t]
  \includegraphics[width=\hsize]{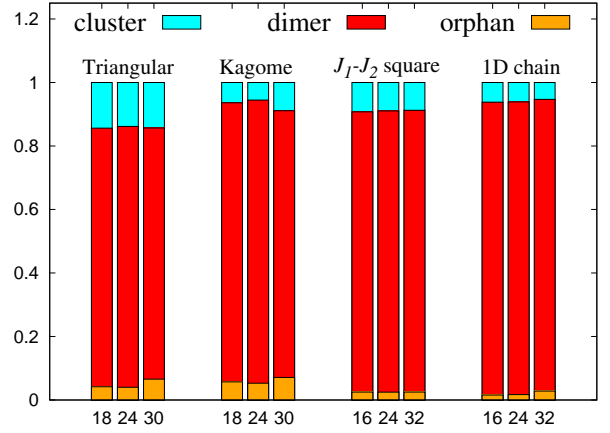}\\
  \caption{\label{fig:order} (Color online)
The ratio of each singlet-dimer configurations, {\it i.e.\/}, isolated singlet-dimers, resonating singlet-dimers clusters and orphan spins, for each of the triangular, the kagome, the $J_1$-$J_2$ ($J_2=0.5$) square, and the 1D lattices of the sizes $N=18$ (triangular and kagome) or 16 ($J_1$-$J_2$ square and 1D), 24, and 30 (triangular and kagome) or 32 ($J_1$-$J_2$ square and 1D).
} 
\end{figure}

 On the basis of the information we have obtained, the nature of the random-singlet state in 2D can be captured as follows. The state primarily consists of nearly isolated singlet-dimers, which are more or less fixed to the bond not resonating with each other. These isolated singlet-dimers are primarily formed between the interaction bonds, {\it i.e.\/}, the NN bonds in the triangular and the kagome lattices while the NN and NNN bonds in the $J_1$-$J_2$ square lattice. Yet, most interesting features of the 2D random-singlet state are the appearance of other types of minor elements, {\it i.e.\/}, a fair amount of {\it resonating singlet-dimers clusters\/} and of {\it orphan spins\/}, each existing with about $\sim $10\% and $\sim $5\% appearance probability. These objects might well be originated from the high frustration effect inherent to the dimer-covering problem on the random 2D lattice. Near energetic degeneracy between more than one distinct dimer-coverings would give rise to resonating singlet-dimers clusters in the quantum model, while the same frustration effect would give rise to orphan spins failed to form well-defined singlet-dimers with neighboring spins.

\section{IV. The nature of low-energy excitations}

 In this section, we study the nature of low-energy excitations in the random-singlet state. For this purpose, we investigate the singlet-dimer configurations of the ground state and of the first excited state of the same sample, and compare them to extract information about the low-energy excitations. As mentioned, the ground state of the model can be either a singlet with $S=0$ or a triplet with $S=1$ in the range of sizes studied, with a rare exception of a quintet with $S=2$. The same is true for the first excited state: It can be either a singlet or a triplet, with a rare exception of a quintet. Then, the lowest-energy excitation can be either singlet-to-singlet, singlet-to-triplet, triplet-to-singlet and triplet-to-triplet excitations. We then examine the character of typical low-energy excitations for each category, to find no clear appreciable distinction in the associated singlet-dimer configurations among them.
 \begin{figure}[t]

  \includegraphics[width=4cm]{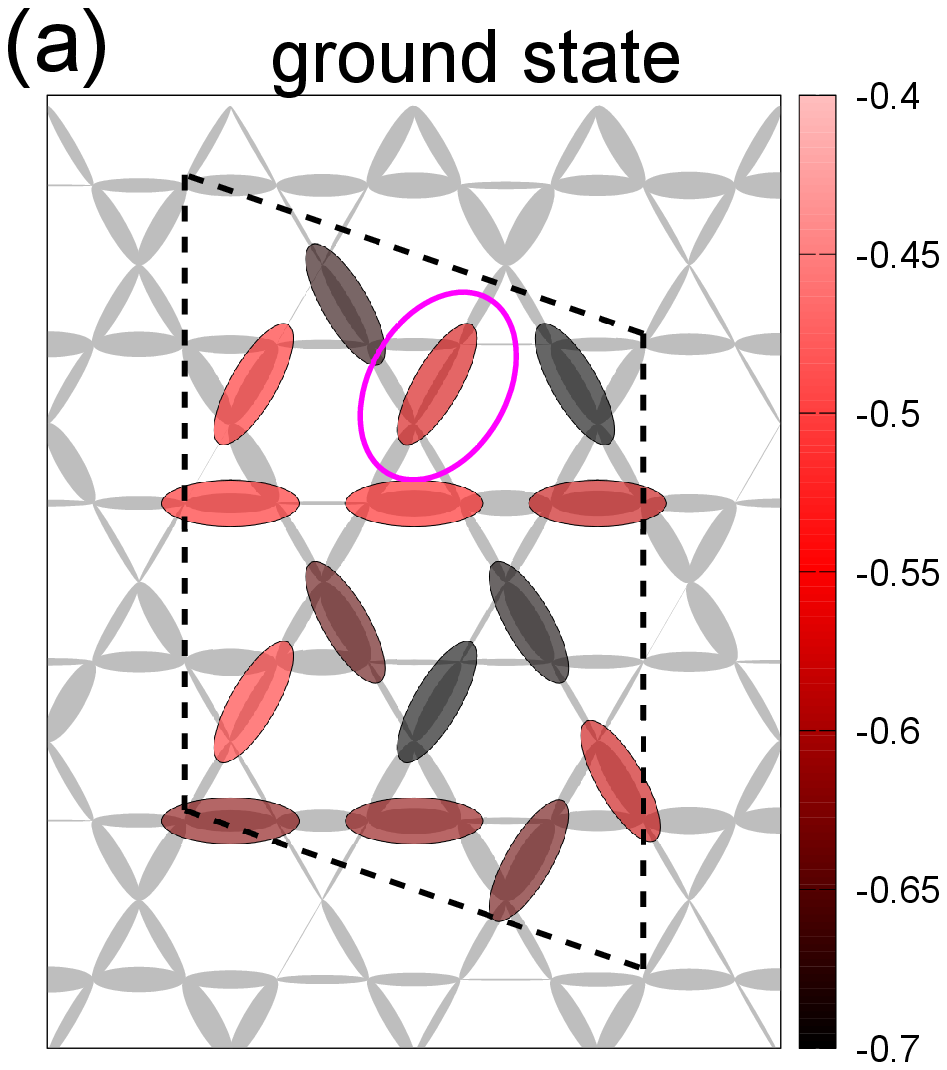}
  \includegraphics[width=4cm]{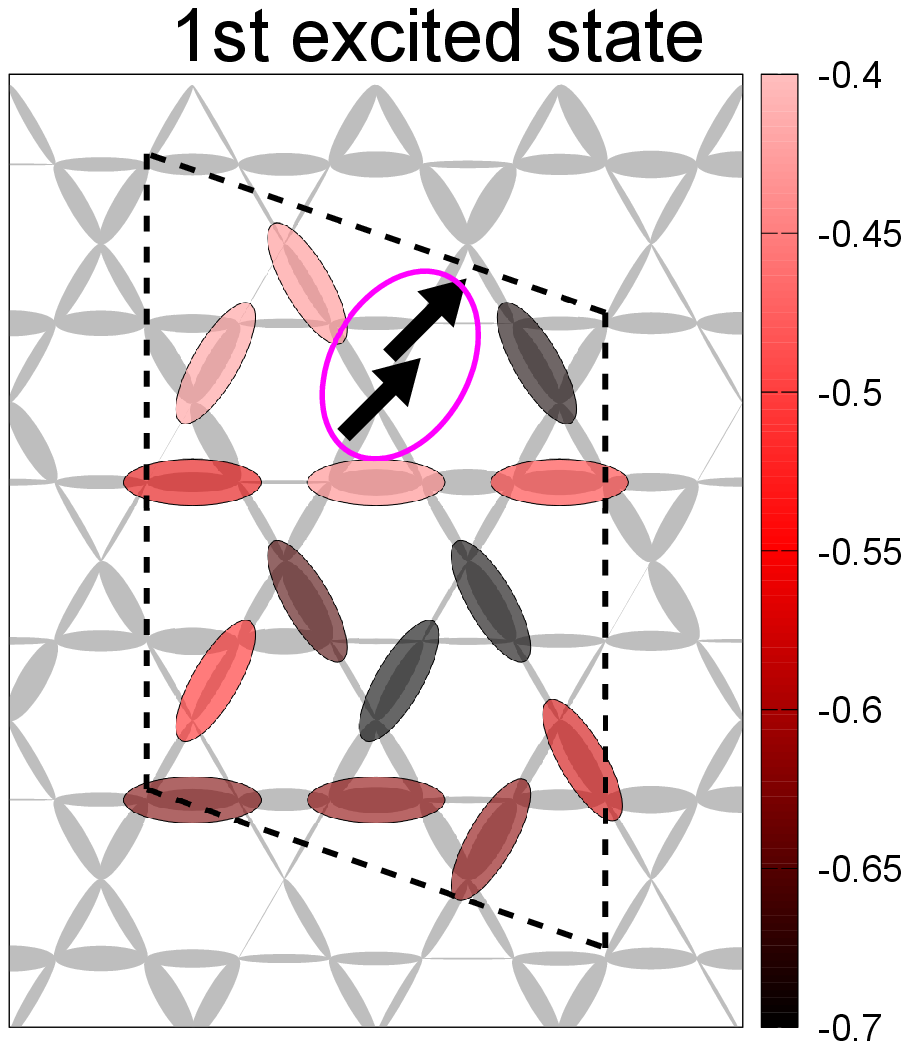}\\
  \includegraphics[width=4cm]{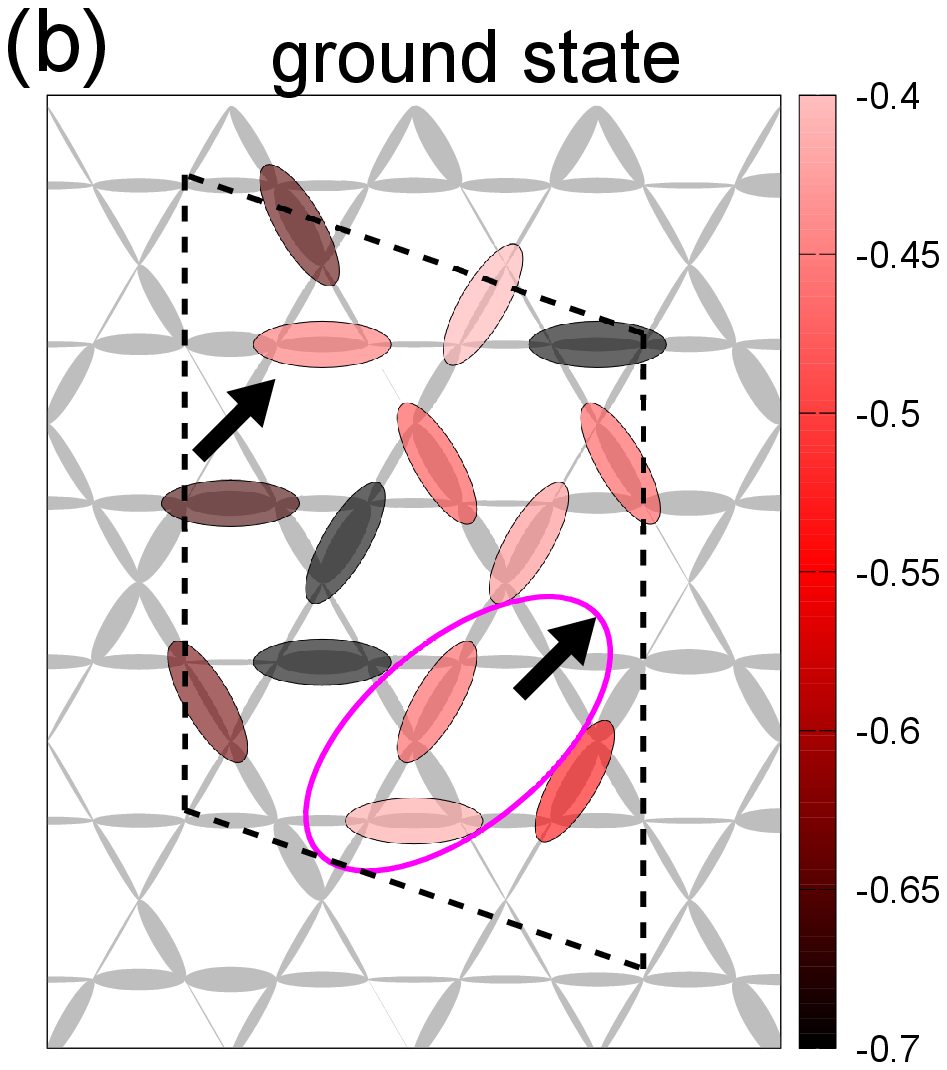}
  \includegraphics[width=4cm]{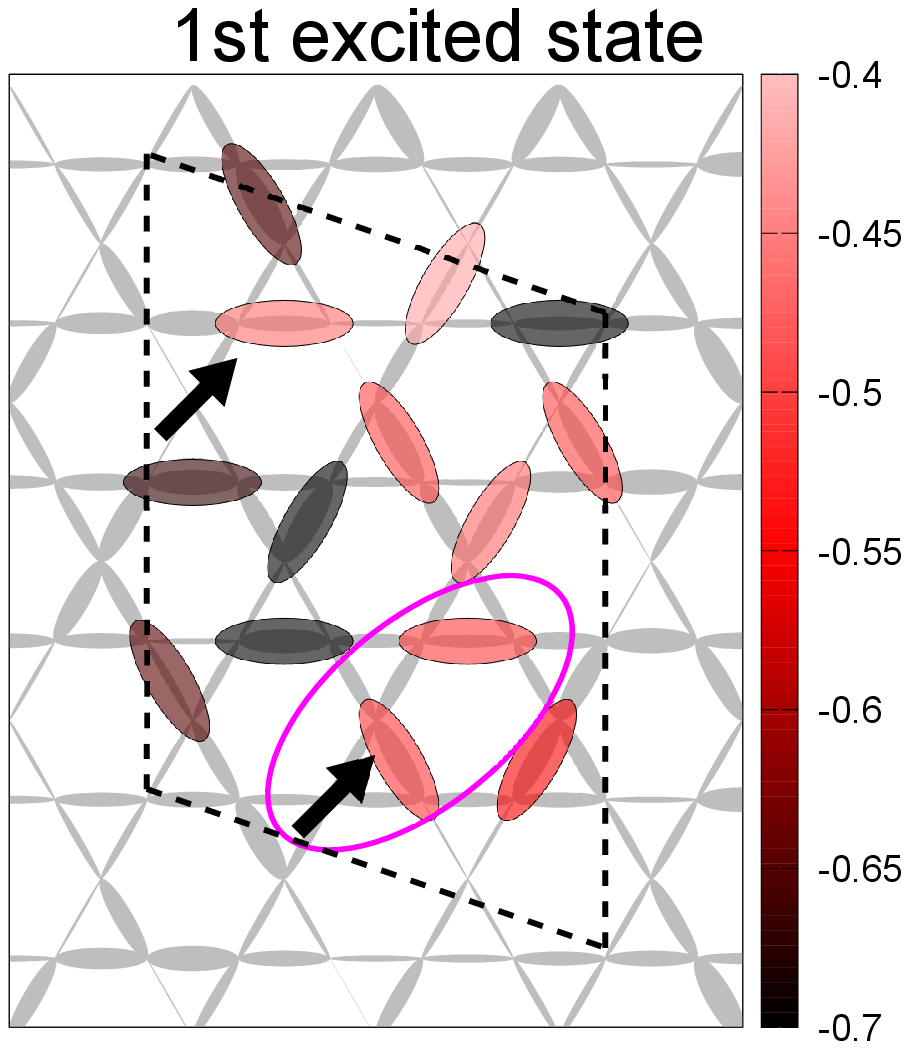}\\
  \includegraphics[width=4cm]{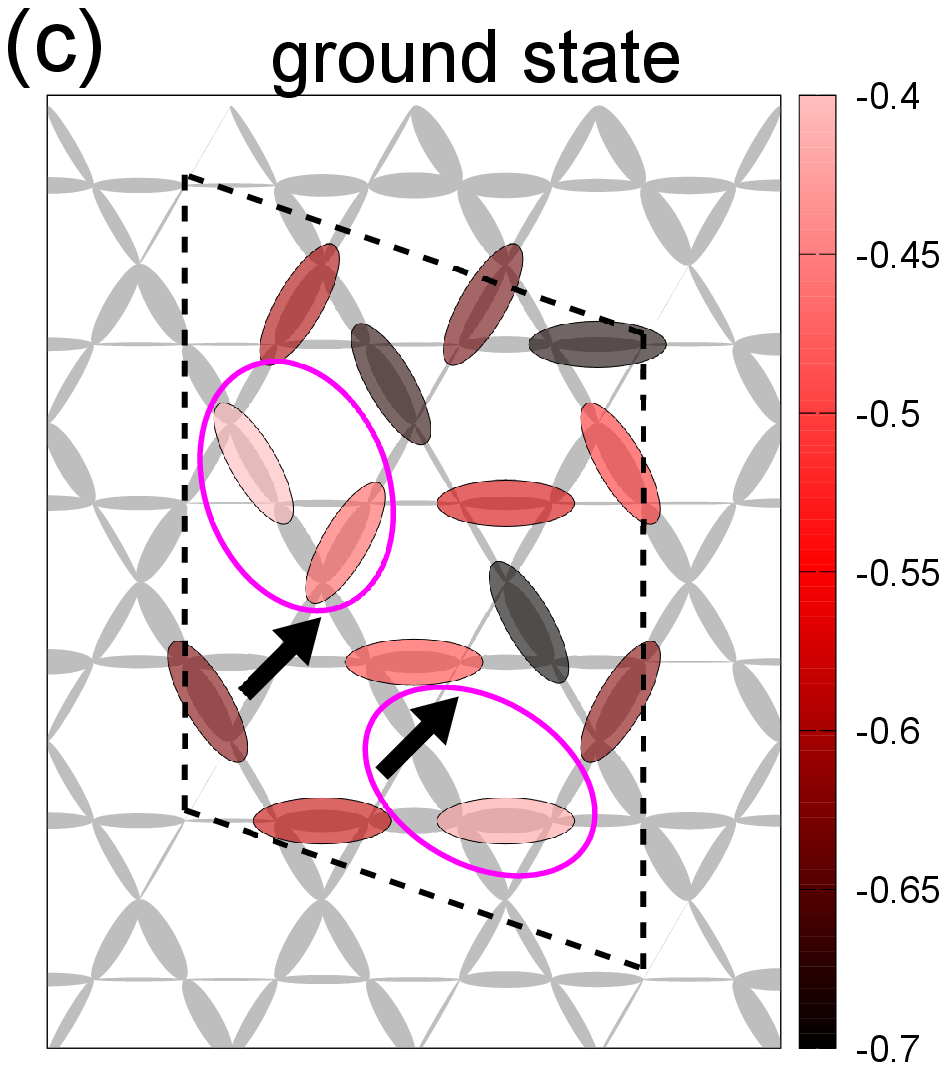}
  \includegraphics[width=4cm]{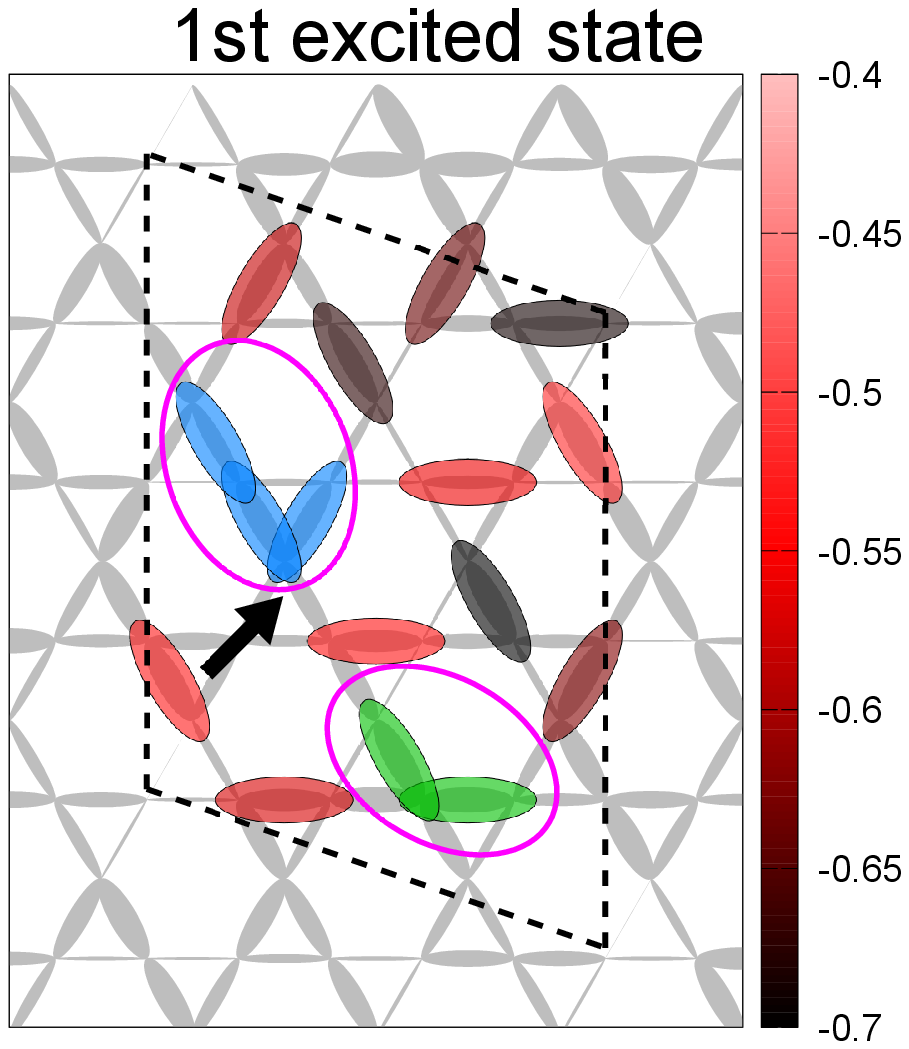}\\
  \caption{
(Color online) 
Typical singlet-dimer configurations of the ground state (left column) and the 1st excited state (the right column) of certain kagome samples of the size $N=30$. Ellipses and arrows represent singlet-dimers and orphan spins, respectively. Gray bonds represent the interaction $J_{ij}$, whose thickness represents its strength. Each low-energy excitation corresponds to (a) breaking of an isolated singlet-dimer into two orphan spins (singlet-to-triplet excitation), (b) diffusion of orphan spins accompanied by the recombination of nearby isolated singlet-dimers, and (c) creation of a cluster of resonating singlet-dimers from isolated singlet-dimers and orphan spin.
} 
\end{figure}

 Then, three distinct types of low-energy excitations, labeled as (A), (B) and (C) here, are identified, {\it i.e.\/}, (A) breaking of an isolated singlet-dimer into two orphan spins (and its reverse process), (B) diffusion of orphan spins accompanied by the recombination of nearby isolated singlet-dimers, and (C) creation (or annihilation) of singlet-dimers clusters from (or into) isolated singlet-dimers (and orphan spin). As a variant of (B), we also observe (B') recombination of nearby isolated singlet-dimers not involving orphan spins, which, however, turns out to be a fairly rare process. Typical examples of these low-energy excitations (A)-(C) are illustrated for the kagome lattice in Figs.6(a)-(c). Of course, the real excited state could be, or generally is a combination of these (A)-(C). For example, the combination of (A) and (B) leads to the type of excitation where a singlet-dimer is gone producing two orphan spins, while orphan spin is not located at its birth-point belonging to the original singlet-dimer, but moves to the neighboring site accompanied by the recombination of singlet-dimers. In fact, in 2D, such a combination type (A)+(B) can be seen quite frequently, while the pure type (A) as shown in Fig.6(a) turns out to be rather rare.

 The type (A) excitation may also be regarded as a local singlet-to-triplet (or triplet-to-singlet) excitation. In fact, we observe that when an isolated singlet-dimer is excited into two orphan spin pair as shown in Fig.6(a), the spin correlation between the two orphan spins tends to be ferromagnetic, $e_{ij}>0$, suggesting the occurrence of the triplet-like state. The type (B) excitations is associated with the diffusive motion of orphan spins. In other words, orphan spins are not entirely localized objects but can ``move'' locally, mimicking a ``spinon'' motion often discussed in the QSL literature. The type (C) excitations are associated with the local resonance of several singlet-dimers. In other words, singlet-dimers are not completely frozen objects, but occasionally exhibit local resonances forming singlet-dimers clusters.

 \begin{figure}[t]
  \includegraphics[width=4.2cm]{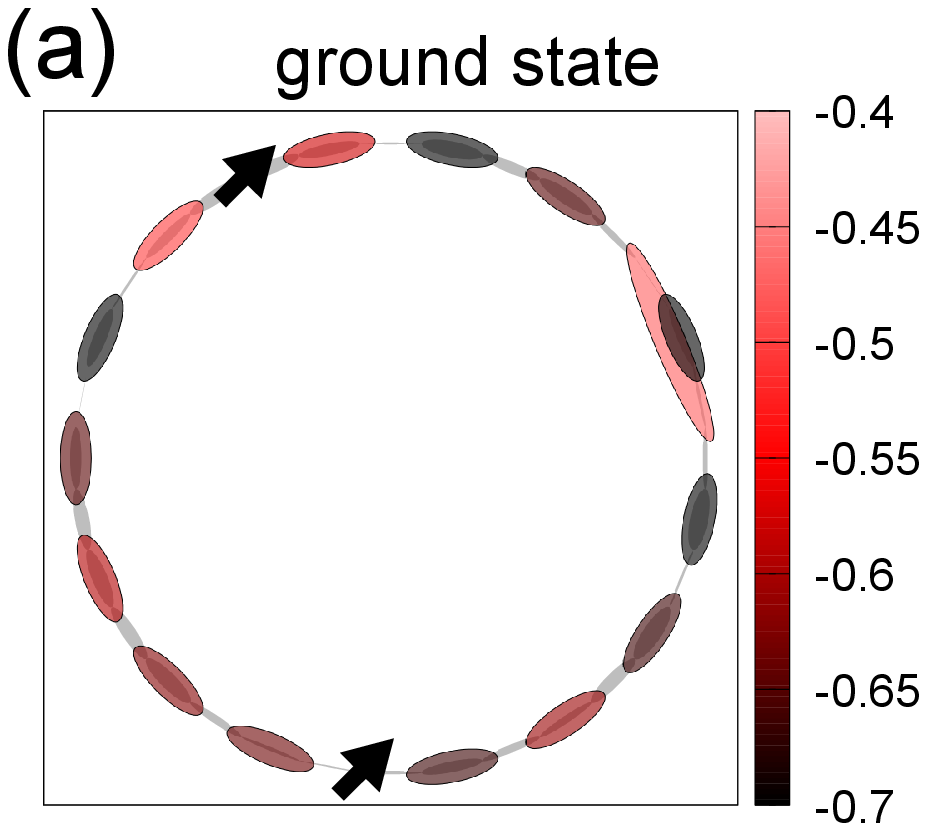}
  \includegraphics[width=4.2cm]{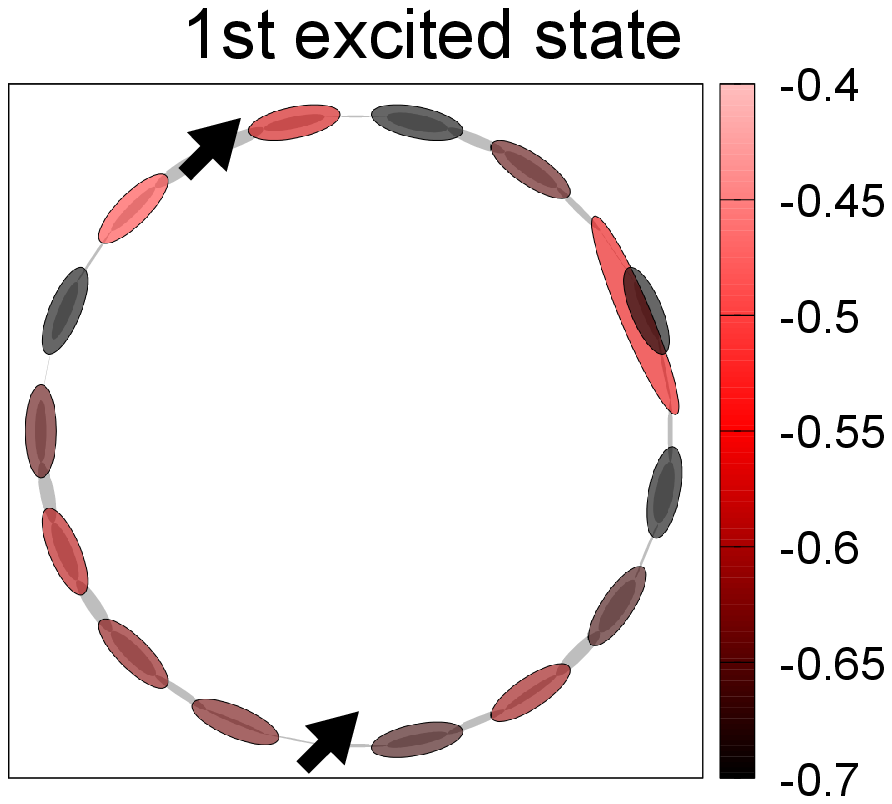}\\
  \includegraphics[width=4.2cm]{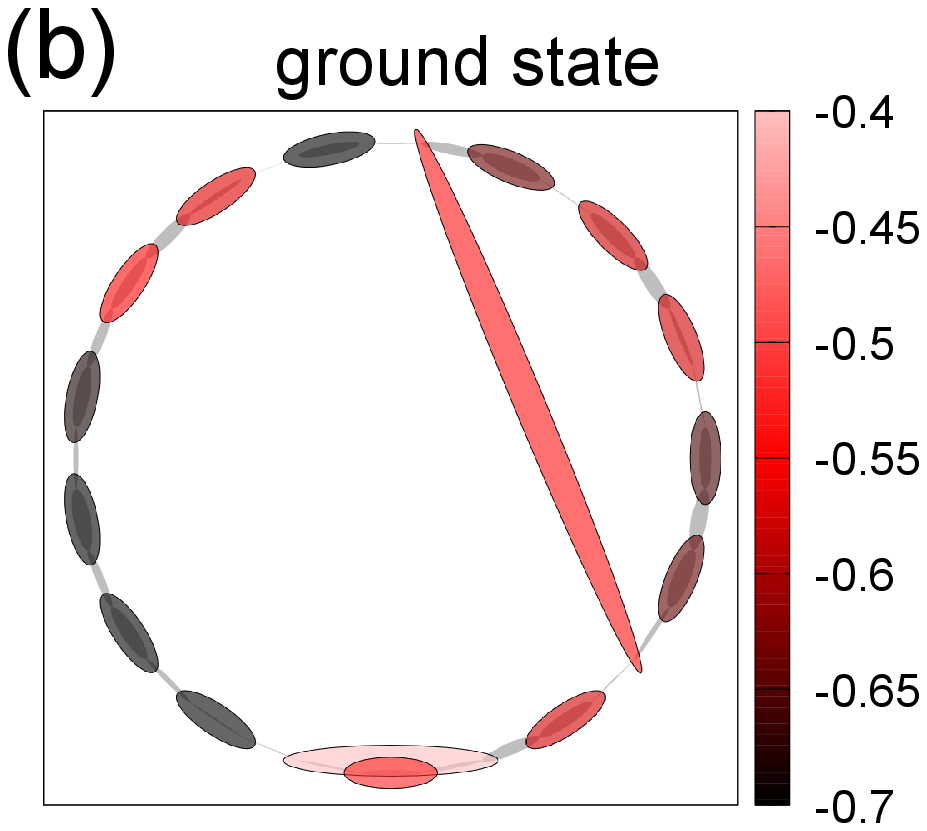}
  \includegraphics[width=4.2cm]{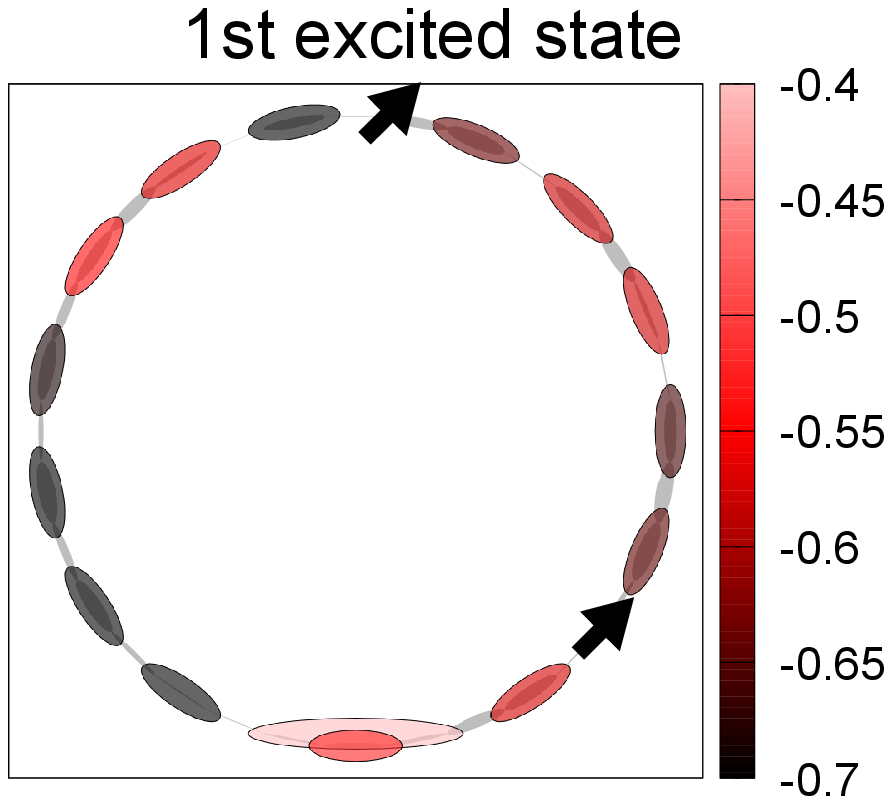}\\
  \caption{\label{fig:order} (Color online) 
Typical singlet-dimer configurations of the ground state (left column) and of the 1st excited state (right column) of certain samples of the 1D NN model of the size $N=32$. Ellipses and arrows represent singlet-dimers and orphan spins, respectively. Gray bonds represent the interaction $J_{ij}$, whose thickness represents its strength. Each low-energy excitation corresponds to (a) excitation not accompanying an appreciable change in the singlet-dimer configuration, and (b) splitting of an isolated singlet-dimer between further-neighbor sites into an orphan spin pair (singlet-to-triplet excitation).
.
} 
\end{figure}

 We also investigate the low-energy excitations of the corresponding 1D NN model. In fact, we find in 1D that the low-energy excitations  exhibit less variety than in 2D studied above. In 1D, the dominant low-energy excitations are of the type different from the ones realized in 2D, {\it i.e.\/}, the type (D), an excitation accompanying no appreciable change in their singlet-dimer configurations from the ground-state. An example is shown in Fig.7(a). Such low-energy excitations are virtually absent in 2D, a few exceptional realizations being observed only for the kagome lattice but no others. Some fraction of low-energy excitations in 1D turn out to be of the type (A), {\it i.e.\/}, an isolated signet-dimer splitting into the orphan spin pair (singlet-to-triplet excitation). Unlike the 1D case, such an orphan spin pair excitation tends to be formed primarily on further-neighbor singlet-dimers which exist with a relatively high rate in 1D. An example is shown in Fig.7(b). In fact, the type (D) excitation is actually quite close to the type (A), {\it i.e.\/}, it is a virtually singlet-to-triplet excitation. In an example Fig.7(a), two distant-neighbor spins, which are just short of forming a singlet-dimer because of the associated AF $e_{ij}$-value, $e_{ij}=-0.24$, being slightly larger than the threshold-value, $e_c=-0.25$, and are counted as two orphan spins, are excited to triplet-forming spin pair with a ferromagnetic correlation of $e_{ij}=0.08$. Anyway, a notable feature of the low-energy excitations in 1D is that there is virtually no type (B) nor (C) excitation.

 Thus, there exist major difference between the random-singlet state in 2D and in 1D concerning their low-energy excitations, rather than concerning their ground states themselves. In the random-singlet state in 1D, the singlet-dimer configurations including the orphan-spin distribution are more or less `frozen' to its $\{ J_{ij} \}$ pattern in the sense that even in the low-lying excited states the singlet-dimer configuration remain basically unchanged, or limited to singlet-to-triplet excitations of distant-neighbor singlet-dimers. The random-singlet state in 2D is more `dynamical' in the sense that the orphan-spin diffusion and the local resonance or clustering of distinct singlet-dimers occur. In this way, the 2D random-singlet state is a more `liquid-like' state than in 1D.

 The origin of the $T$-linear specific heat generically observed in the 2D random-singlet state (see also the next section) has been ascribed to the nonzero weight of the density of states of low-energy excitations, $\rho(E \rightarrow 0)>0$ \cite{Uematsu2}. According to the phenomenological argument of Ref.\cite{AndHalVar}, these low-energy excitations in the random system are associated with the local cluster-type excitations (two-level system) arising from their random environments. Our present analysis suggests that the local excitations with a near-constant density of states postulated in such a phenomenological picture are actually of the types (A)-(C), {\it i.e.\/},  an isolated singlet-dimer breaking into two orphan spins (and its reverse process), diffusion of orphan spins accompanied with the recombination of nearby isolated singlet-dimers, creation (or annihilation) of singlet-dimers clusters from (or into) isolated singlet-dimers (and orphan spin). These excitations are expected to bear the $T$-linear low-$T$ specific heat. In 1D, by contrast, the earlier theoretical analysis suggested a different asymptotic form for the low-$T$ specific heat, {\it i.e.\/}, $\sim 1/|\log T|^3$ \cite{Hirsch}. %Our present analysis suggests that the excitation bearing this behavior might of the type (D) one, {\it i.e.\/}, the excitations not accompanying any change in the singlet-dimer configuration: See also the next section.

\section{V. Finite-temperature properties}

 In this section, we study thermodynamic properties of the model at finite temperatures, more specifically, the specific heat and the susceptibility. In 2D, these quantities were already studied for each respective model in Refs.\cite{Watanabe,Kawamura,Uematsu2} at low temperatures of, say, $T\lesssim 0.2$. Here, we re-examine these quantities in a wider temperature range of  $T\lesssim 2$, comparatively among various 2D models, and also with the one of the $J_1$-model in 1D.

 \begin{figure}[t]
  \includegraphics[width=6cm]{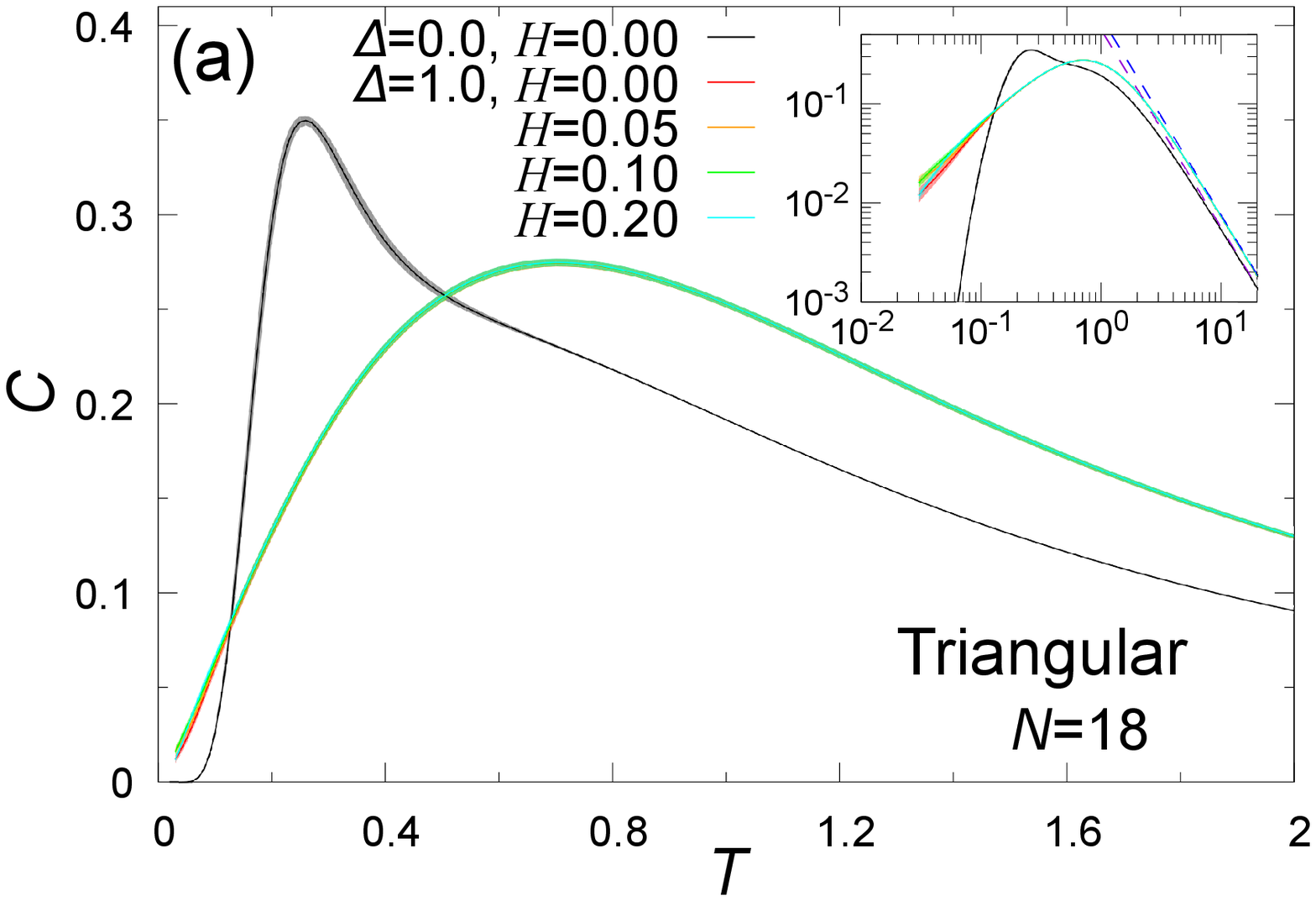}\\
  \includegraphics[width=6cm]{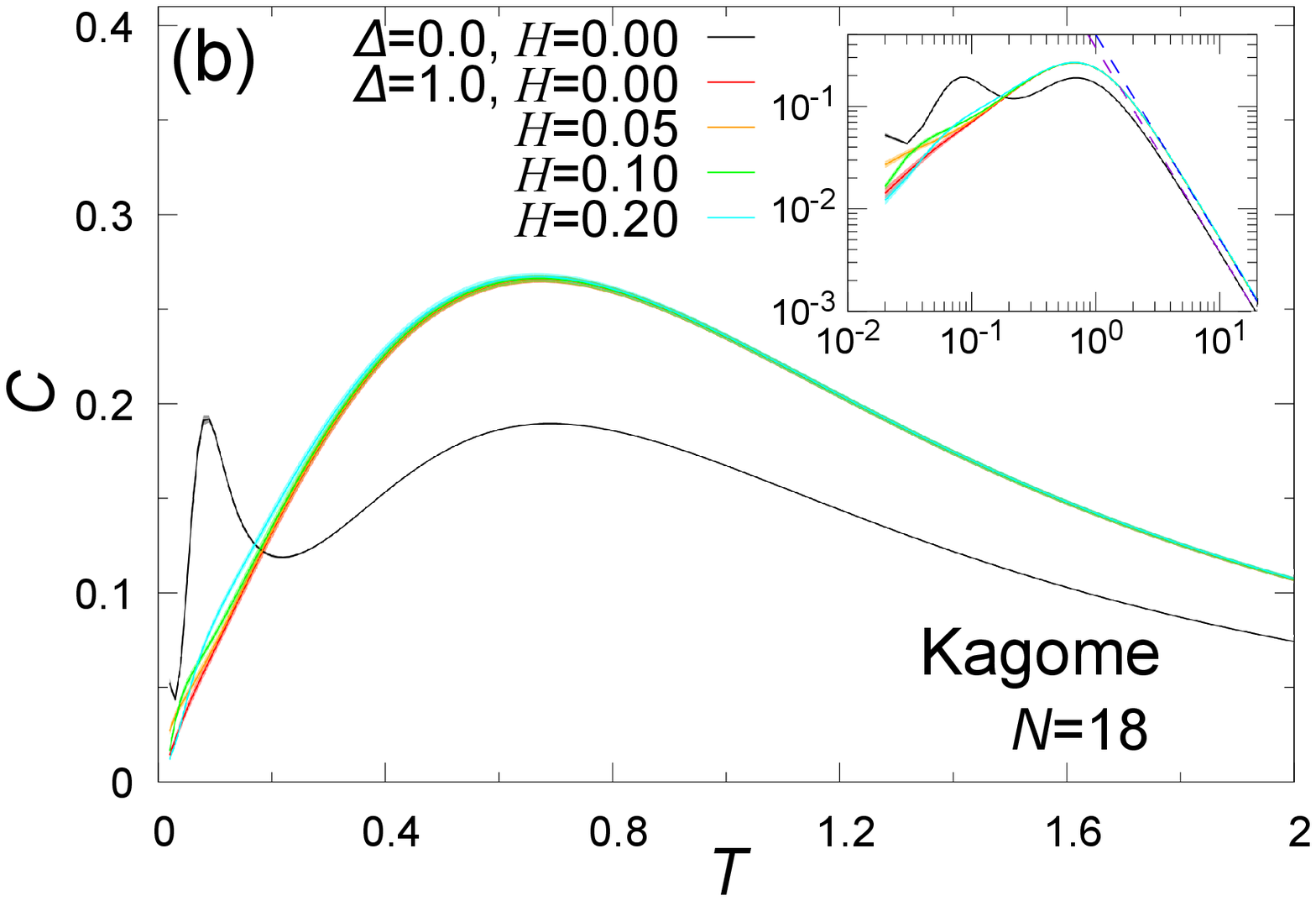}\\
  \includegraphics[width=6cm]{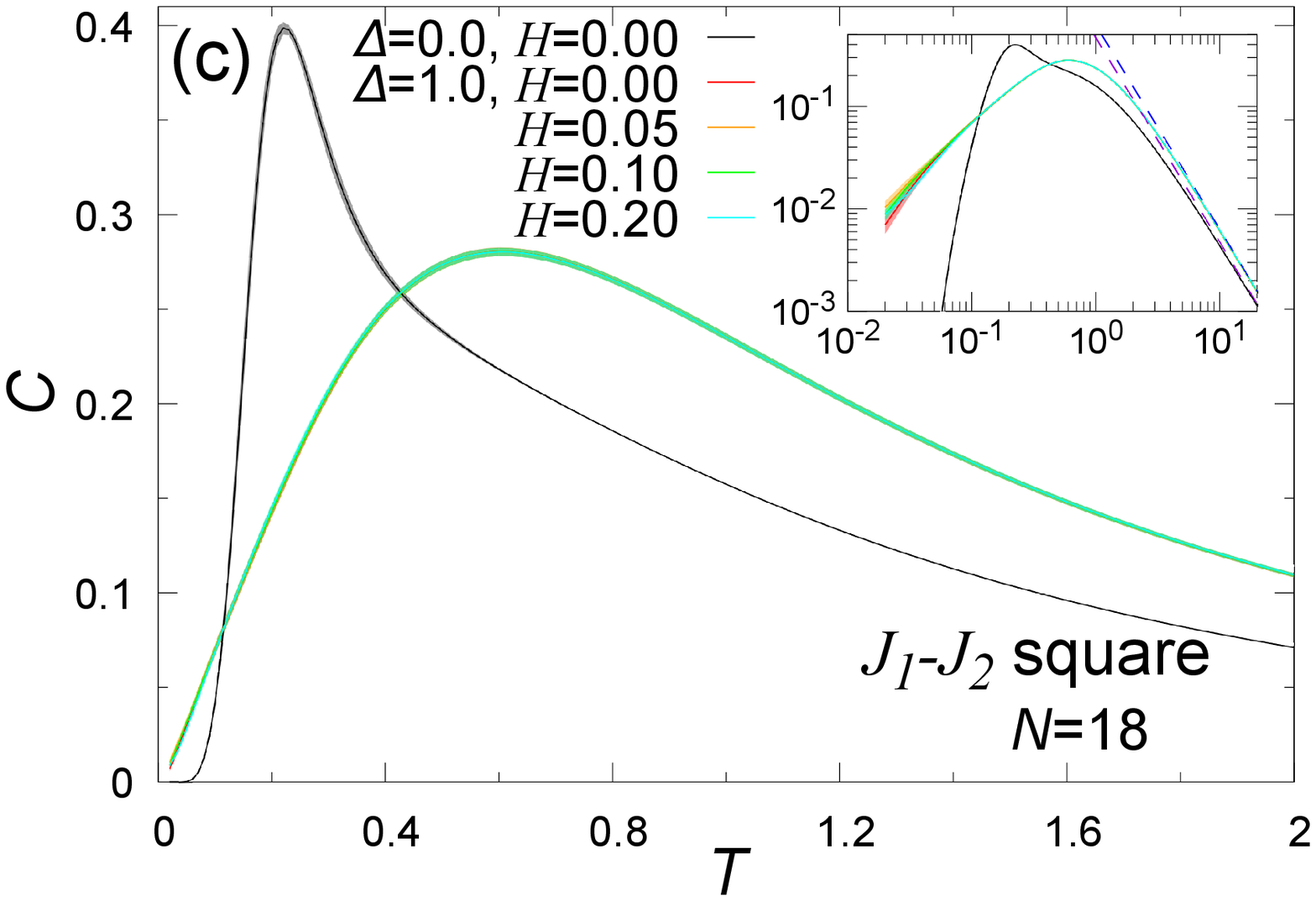}\\
  \includegraphics[width=6cm]{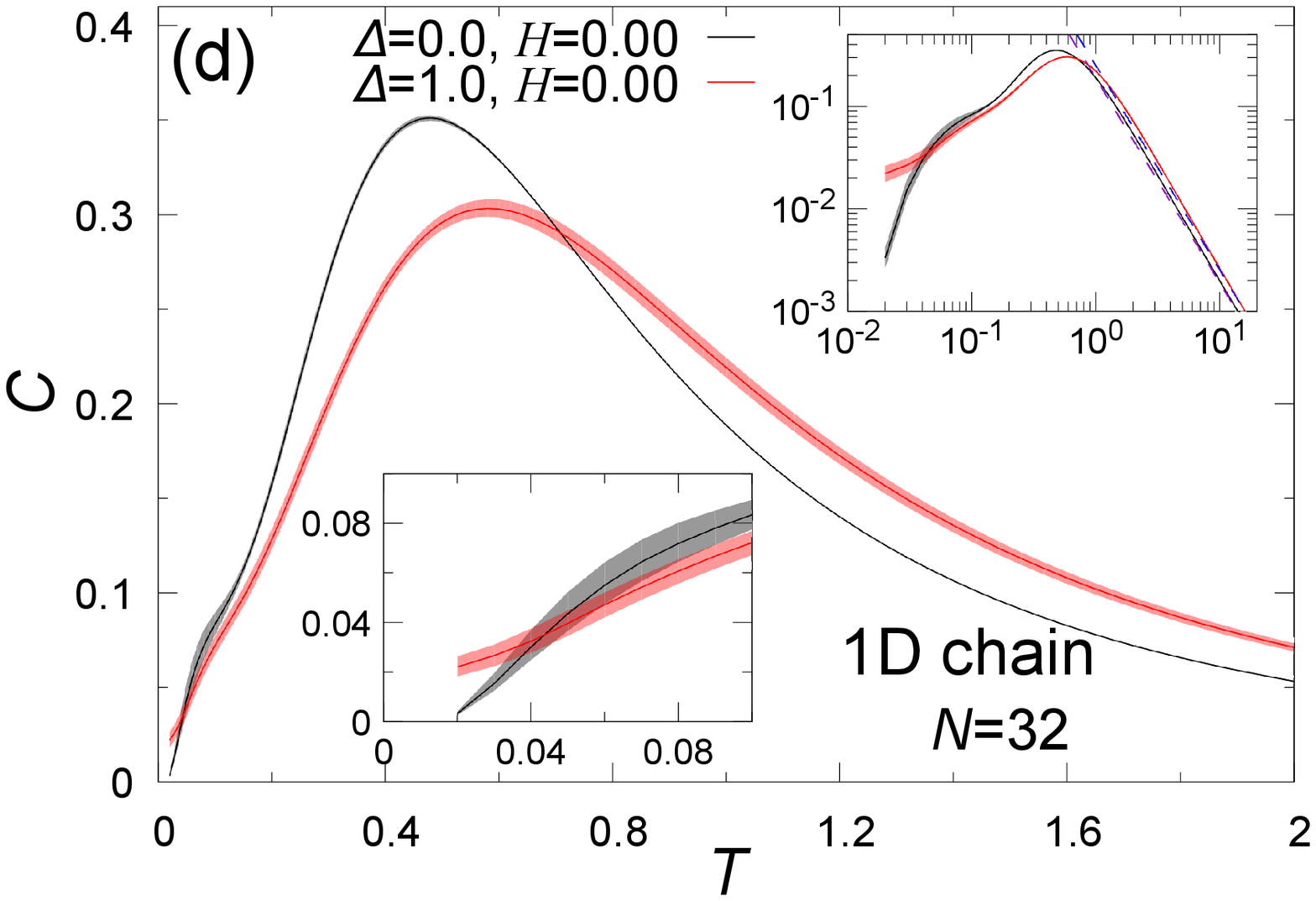}\\
  \caption{\label{fig:order} (Color online) 
The temperature dependence of the specific heat of the random model $\Delta=1$ in zero and weak fields for (a) the triangular, (b) the kagome, (c) the $J_1$-$J_2$ ($J_2=0.5$) square, and (d) the 1D lattices, shown with the one of the regular model $\Delta=0$. Insets represent the wider temperature range $T\leq 20$, while the lower inset in (d)  the lower-$T$ range. The dashed lines are the high-$T$ expansion results. The system size is $N=18$ (2D) and $N=32$ (1D).
} 
\end{figure}

 In Fig.8, we show the temperature dependence of the specific heat per spin (measured in units of $k_B$) in the temperature range of $T\leq 2$ of the regular model of $\Delta=0$ and of the maximally random model of $\Delta=1$, for the cases of (a) the triangular, (b) the kagome, (c) the $J_1$-$J_2$ ($J_2=0.5$) square, and (d) the 1D lattices. Double-logarithmic plots in the wider temperature regime of $T\leq20$ are given in the insets. The lattice sizes are $N=18$ in 2D (the system-size dependence was studied in Refs.\cite{Watanabe,Kawamura,Uematsu2} in 2D), and $N=32$ in 1D. In the random-singlet state in 2D, the specific heat exhibits a $T$-linear behavior as already identified in earlier works, as can be seen from Figs.8(a)-(c).  The data for certain intermediate values of $\Delta$ have been given in Refs.\cite{Watanabe,Kawamura,Uematsu,Uematsu2}.

 In 2D models, we also compute the specific heat under weak fields of $H\leq 0.2$, where the field intensity $H$ is measured in units of $J_1$. (The magnetic field contributes to the Hamiltonian (1) via the Zeeman term $-H\sum_i S_i^z$ where we have taken the $g\mu_B=1$ unit.) As can be seen from Figs.8(a)-(c), the low-$T$ specific heat is insensitive to applied weak fields, although a weak field dependence is discernible in the low-$T$ region of the kagome case. The observed field insensitivity is consistent with the property observed in Ref.\cite{Watanabe} for the triangular model. Such a field insensitivity of the low-$T$ specific heat, especially of the $T$-linear part, also agrees with the experimental observation on triangular organic salts $\kappa$-(ET)$_2$Cu$_2$(CN)$_3$ \cite{ET-Nakazawa} and EtMe$_3$Sb[Pd(dmit)$_2$]$_2$ \cite{dmit-Nakazawa}. For the square-lattice magnet Sr$_2$Cu(Te$_{1-x}$W$_{x}$)O$_6$, it  was also reported that the $T$-linear part of the low-$T$ specific heat was insensitive to the applied field \cite{Tanaka}. For the kagome herbertsmithite, the low-$T$ specific heat seems to exhibit a weak field dependence even in low fields \cite{Helton}, though one should notice that the situation in herbersmithite is further complicated due to the contribution from the Cu$^{2+}$ spins randomly substituting Zn$^{2+}$ on the sparse triangular layer between the kagome layers. 

 A slightly different behavior is observed for the 1D NN chain, for which the earlier theoretical analysis predicted the $1/|\log T|^3$ behavior \cite{Hirsch}. As can be seen from the inset of Fig.8(d), a linear extrapolation of the low-$T$ data ($0.02\leq T\leq 0.04$) yields a nonzero tangent in the $T\rightarrow 0$ limit, suggesting the existence of an appreciable positive curvature in the lower temperatures range, which is consistent with the expected logarithmic behavior \cite{Hirsch}.

 At higher temperatures, the behavior of the specific heat of the random models is qualitatively the same between in 2D and in 1D. They all exhibit a single rounded peak at a temperature of order $J$ (a bit low than that) reflecting the absence of any characteristic energy scale other than the interaction energy $J$ in the excitation-energy spectrum of the state. At still higher temperatures of $T\gtrsim 2$, the specific heat of both the regular and the random models decays as $1/T^2$ with different coefficients. The behavior at higher temperatures can be well understood from the lowest-order high-$T$ expansion results shown in the insets of Figs.8(a)-(d) by dashed lines both for the random (upper line) and regular (lower line) models.

 \begin{figure}[t]
  \includegraphics[width=6cm]{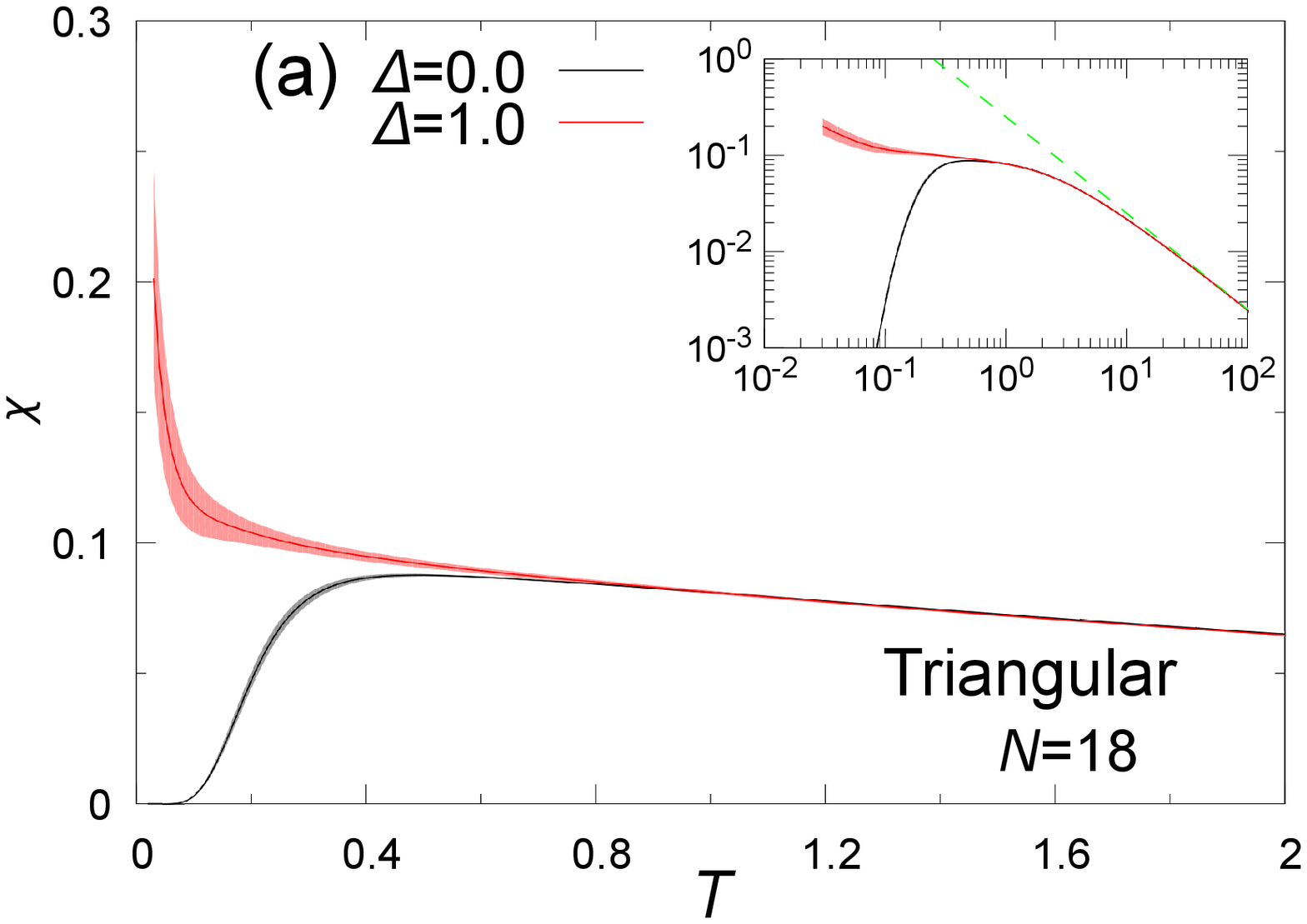}\\
  \includegraphics[width=6cm]{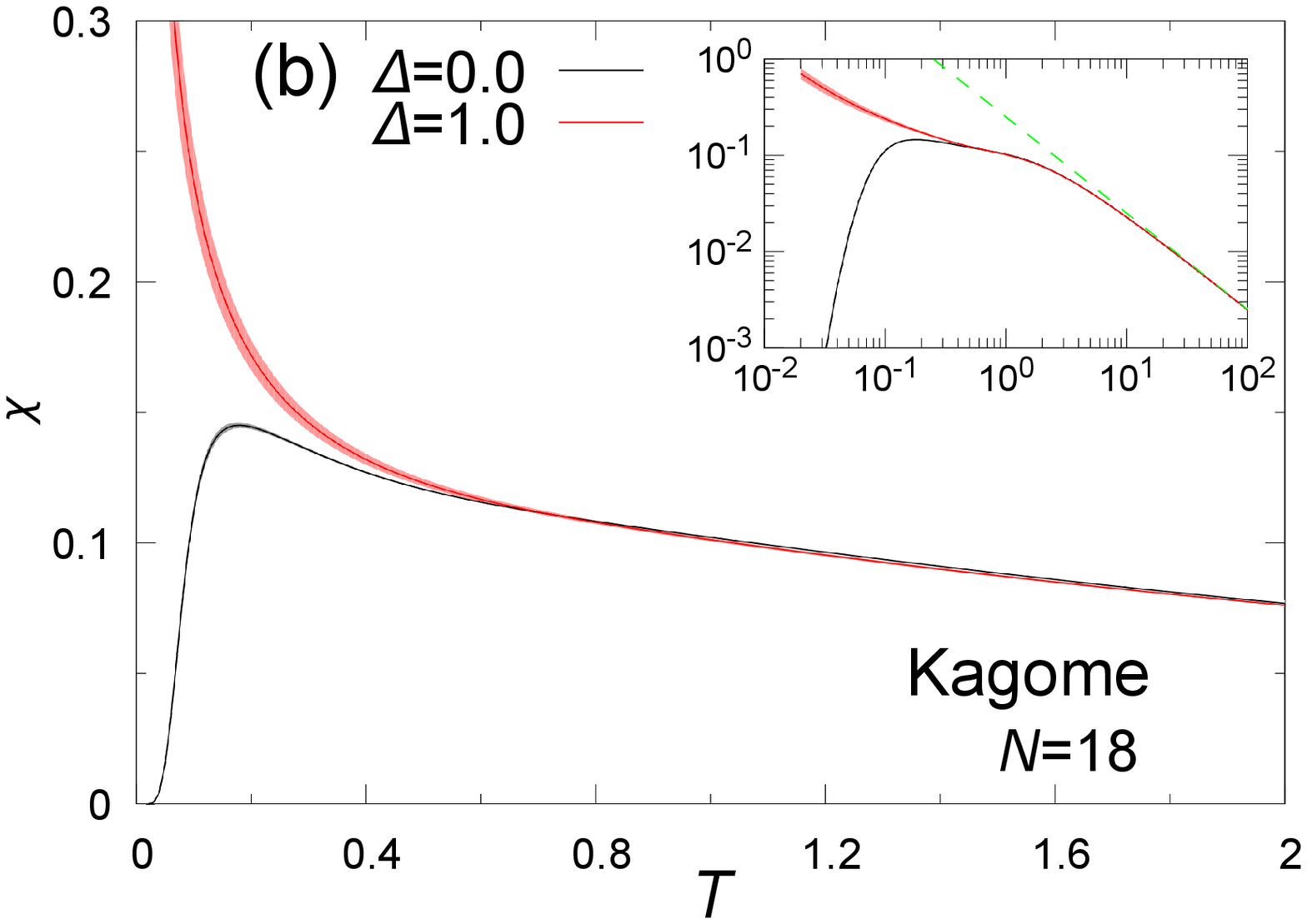}\\
  \includegraphics[width=6cm]{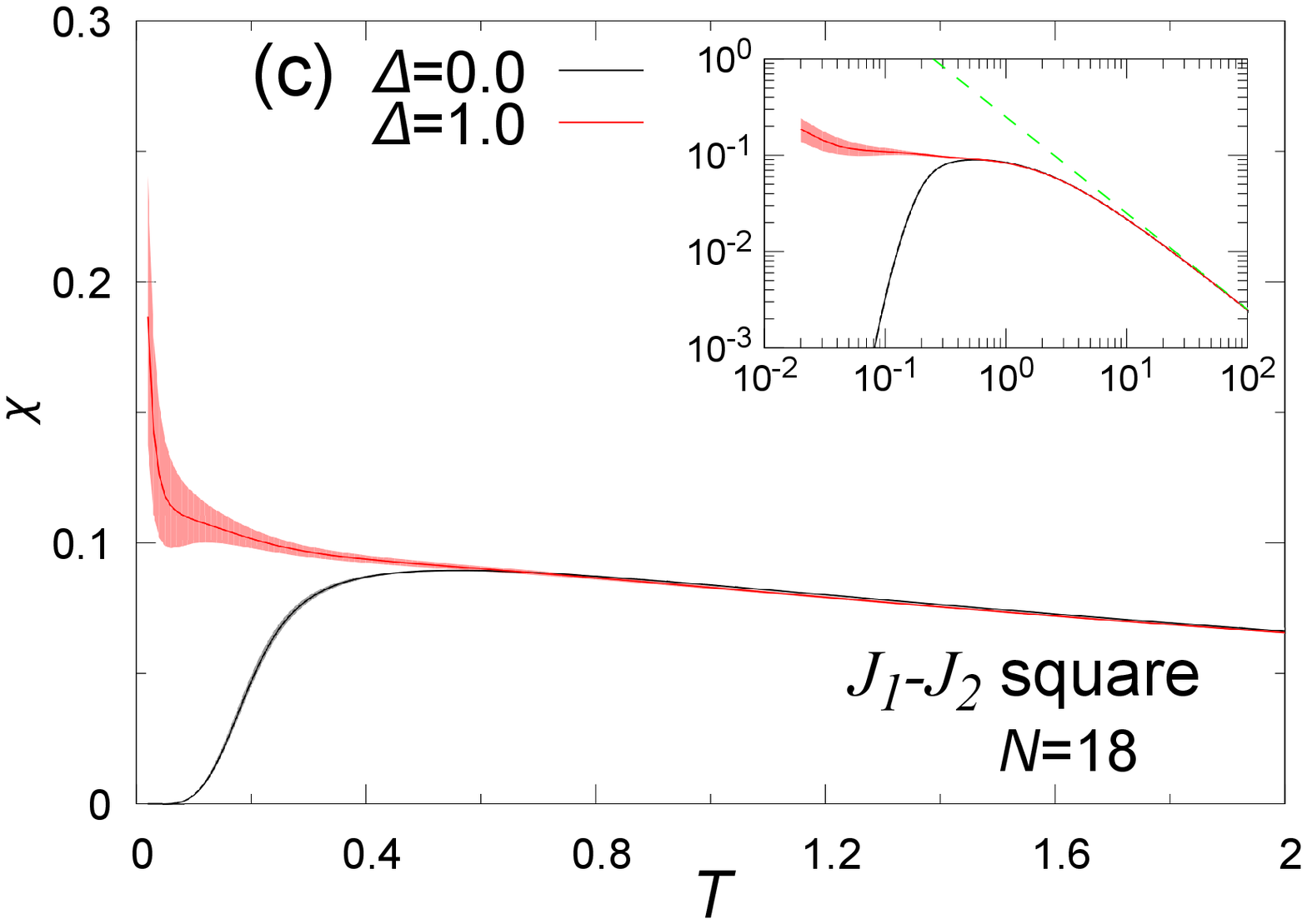}\\
  \includegraphics[width=6cm]{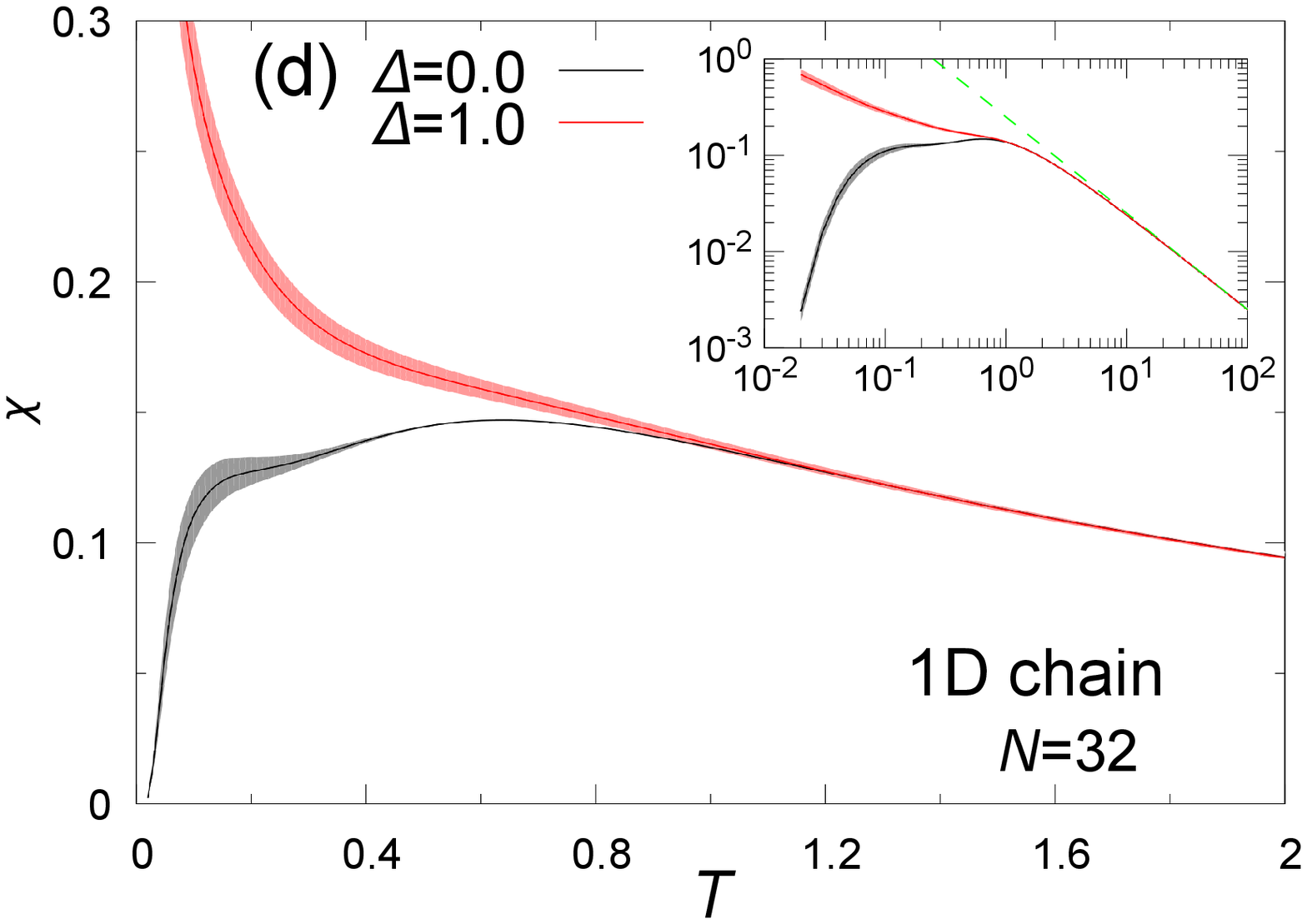}\\
  \caption{\label{fig:order} (Color online) 
The temperature dependence of the susceptibility of the random model $\Delta=1$ for (a) the triangular, (b) the kagome, (c) the $J_1$-$J_2$ ($J_2=0.5$) square, and (d) the 1D lattices, shown with the one of the regular model $\Delta=0$. Insets are double-logarithmic plots in a wider temperature range $T\leq 100$. The dashed lines are the high-$T$ expansion results. The system size is $N=18$ (2D) and $N=32$ (1D).
} 
\end{figure}

 In Fig.9, we show the temperature dependence of the uniform susceptibility per spin of the regular model of $\Delta=0$ and of the maximally random model of $\Delta=1$, for the cases of (a) the triangular, (b) the kagome, (c) the $J_1$-$J_2$  ($J_2=0.5$) square, and (d) the 1D lattices in the same temperature range of $T\leq 2$ as the specific heat. (Note that we have taken the $g\mu_B=1$ unit here.)  Double-logarithmic plots in the wider temperature regime of $T\leq 100$ are given in the insets. The system size is $N=18$ in 2D, and $N=32$ in 1D. The data for certain intermediate values of $\Delta$ have been given in Refs.\cite{Watanabe,Kawamura,Uematsu,Uematsu2}.

 For the susceptibility, no appreciable difference is observed between in 2D and in 1D even in the low-$T$ regime where all the models turn out to exhibit a Curie-like tail. Indeed, for 1D, the earlier theoretical analysis predicted the behavior $\chi \sim 1/(T|\log T|^2)$ \cite{Hirsch,Fisher}, which, however, is practically indistinguishable from the standard Curie law for small systems as studied here. As can clearly be seen from the figure, the susceptibility tends to take a nearly temperature-independent value of $\sim 0.1$ in the intermediate temperature range of $0.2\lesssim T\lesssim 2$, while it exhibits a $1/T$ behavior corresponding to the standard Curie law at higher-temperature of $T\gtrsim 2$. These Curie-law behavior at higher temperatures can be well understood from the lowest-order high-$T$ expansion results shown in the insets of Figs.9(a)-(d) by dashed lines which are identical in the random and the regular cases.

 Meanwhile, at lower temperature range of $T\lesssim 0.2$, it exhibits another Curie-like law as noted above, probably arising from the orphan spins. In fact, since orphan spins are not quite free spins, it is not entirely clear whether this Curie-like tail at low temperatures extends down to $T\rightarrow 0$, and if it is to really diverge at $T\rightarrow 0$, whether the associated exponent is just unity or not, especially in the 2D case for which the meaningful RG analysis of the random-singlet state is absent at present. It is difficult to clarify this issue from the ED calculation, and we need more powerful numerical approach to probe larger lattices and lower temperatures.

 In contrast to the specific heat, the susceptibilities of the regular and the random models coincide in the temperature range $T\gtrsim 0.2$. In the high-$T$ Curie-law regime of $T\gtrsim 2$, such coincidence is just as expected from the high-temperature expansion, whereas, in the near-constant $\chi$ regime of $0.2\lesssim T\lesssim 2$, the coincidence seems highly nontrivial. One may think that the randomness is irrelevant in this intermediate temperature range, but the specific heats  exhibit entirely different behavior between in the regular and the random models in exactly the same temperature range.

\section{VI. Summary and discussion}

\noindent {\it Summary of the present results}

The nature of the randomness-induced quantum spin liquid state, the random-singlet state, is investigated in two dimensions (2D) by means of the ED and the Hams-de Raedt methods. We study several random-bond $s=1/2$ frustrated Heisenberg models sustaining the random-singlet state, including the triangular, the kagome and the $J_1$-$J_2$ square lattice models. The properties of the ground state, the low-energy excitations and the finite-temperature thermodynamic properties are investigated. Comparison is made with the random-bond $s=1/2$ unfrusrated Heisenberg model in 1D, with the hope to get deeper insight into the random-singlet state in 2D by clarifying the similarity to and the difference from the random-singlet state in 1D. We also extend our previous calculations of the thermodynamic properties such as the specific heat and the susceptibility at finite temperatures to the wider temperature range, and make comparison among different 2D models and the 1D model.  By so doing, we wish to establish the qualitative picture of the random-singlet state of 2D frustrated magnets.

 It turns out that the ground state and the low-lying excited states in 2D consists of nearly isolated singlet-dimers, clusters of resonating singlet-dimers, and orphan spins. Low-energy excitations are either splitting of isolated singlet-dimers into orphan spins (singlet-to-triplet excitations) and its inverse process, diffusion of orphan spins accompanied by the recombination of nearby singlet-dimers, creation or destruction of resonating singlet-dimers clusters. The latter two excitations give enhanced dynamical `liquid-like' features to the 2D random-singlet state. We emphasize that in causing such enhanced dynamical features of the 2D random singlet state, frustration in a wider sense, {\it i.e.\/} the one including the competition of distinct types of interactions, plays a vital role. Especially, the frustration effect associated with the dimer-covering problem on the 2D random lattice combined with strong quantum fluctuations would give rise to the enhanced dynamical features in the 2D random-singlet state. In this sense, the 2D random-singlet state of our target here should be called `{\it frustrated random-singlet state\/}'. By contrast, the well-studied random-singlet state in the unfrustrated 1D chain tends to be more `static' in the sense that most of its low-energy excitations are limited to singlet-to-triplet excitations on distant-neighbor sites, not accompanying the change in the singlet-dimer configuration. In other words, singlet-dimers and orphan spins tend to be static in the unfrustrated random-singlet state in 1D, while in 2D they are more dynamical even in the low-energy sector assigning enhanced dynamical `liquid-like' features to the 2D random-singlet state. In this sense, one may call the state as an ``many-body localized RVB state''. The state is hierarchically organized from stronger singlet-dimers to weaker singlet-dimers, and eventually to orphan spins.

\medskip
\noindent {\it Discussion of the present results}

 It has been known that the random-singlet state in 1D is described by an infinite-randomness fixed point \cite{Fisher,Motrunich}. An interesting question is then what is the situation in 2D. Although small lattice sizes of our present ED calculation prevents to say anything definitive about the issue, the observations of Refs.\cite{Watanabe,Kawamura,Shimokawa,Uematsu,Uematsu2}, especially the dynamical `liquid-like' features observed in the present study, does not suggest an infinite-randomness fixed point. The same view was also given by Liu {\it et al\/} on the basis of their QMC data on the square-lattice Heisenberg model with the six-body interaction for much larger systems \cite{Guo}. The random-singlet state in 2D is likely to be governed by {\it a finite-randomness fixed point\/}. Technically, it would mean that the strong-disorder RG, which is expected to be asymptotically exact at an infinite-randomness fixed point, would not be so reliable. It may explain the reason, at least partially, why the strong-disorder RG failed to identify the random-singlet-like fixed point in 2D \cite{Lin}.

 Qualitative features of the 2D random-singlet state in the strong random case \cite{Watanabe,Kawamura,Shimokawa,Uematsu,Uematsu2} look very much similar to those of the weak random case of Refs.\cite{Kimchi,Kimchi2,Guo}. So, a natural possibility is be that a randomness-induced QSL state, the 2D random-singlet state, is governed by a unique finite-randomness fixed point, which governs the long-scale physics of both strong and weak random systems, even though some of the short-scale properties look very different. Of course, we cannot rule out a possibility that more than one randomness-induced QSL fixed points exist in 2D, say, the weak and the strong random-singlet fixed points. If so, long-scale properties would differ between the weak and the strong random systems. At the present stage, however, we do not have any clear indication of such distinct random-singlet-like states in 2D. In this connection, the square-lattice model analyzed in Ref.\cite{Guo}, which is a frustrated system in our criterion, is expected to belong to the same class as the frustrated systems analyzed here and in Refs.\cite{Watanabe,Kawamura,Shimokawa,Uematsu,Uematsu2}. An obvious advantage of the model treated in Ref.\cite{Guo} is that it is amenable to QMC, which enables to treat larger systems. Indeed, Liu {\it et al\/} estimated the dynamical exponent $z\geq 2$ \cite{Guo}, which certainly supports the finite-randomness fixed point as opposed to the infinite-randomness one where $z=\infty$ \cite{Fisher,Motrunich}.

 % If the model analyzed in Ref.\cite{Guo} is really to belong to the same universality class as the models analyzed in Refs.\cite{Watanabe,Kawamura,Shimokawa,Uematsu,Uematsu2} and in the present paper, the associated criticalities are expected to be common. The QMC analysis of Ref.\cite{Guo} yields the susceptibility exhibiting the asymptotic behavior of $\chi \sim T^{(D/z)-1}=T^{(2/z)-1}$ with $z\geq 2$. Thus, the asymptotic susceptibility exponent might well be smaller than unity, and the ``Curie-like tail'' might not be the true Curie tail asymptotically. Even a non-divergence cannot be ruled out if $z=2$. Further study is required to quantify the asymptotic critical behavior.

% The present picture of the random-singlet state in the strong random case is more or less dynamical in that the local resonance of singlet-dimers leading to the formation and destruction of resonating singlet-dimers clusters, or the diffusion of orphan spins accompanied by the recombination of nearby singlet-dimers, take place. Although the starting picture of the random-singlet state of Refs.\cite{Kimchi,Guo} in the weak random case seems to be more `static', the dynamical feature on longer-length scale, {\it eg.\/}, the motion of ``spinons'' along the VBC domain wall, was also pointed out in Ref.\cite{Guo}, suggesting that the random-singlet states in the strong and the weak random cases are essentially the same state eventually. 

 The dynamical features of the 2D random-singlet state bring about some interesting consequences. Kimchi {\it et al\/} \cite{Kimchi} and Liu {\it et al\/} \cite{Guo} suggested that, in the weak random case, the 2D random-singlet state eventually became unstable against the spin-glass order on longer length scales. If so, the QSL-like behavior could appear only at intermediate length scales. In apparent contrast, many QSL materials so far identified in 2D and 3D frustrated magnets do not exhibit any SG freezing down to low temperatures in spite of their gapless features. The argument of Refs.\cite{Kimchi,Guo} was based on the strong-disorder RG result \cite{Westerberg,Westerberg2} applied to the spin-1/2 carrying defect spin or the spinon. The implicit assumption there was the defect spins or the spinons were fixed in space, not changing their positions. In contrast to this, the orphan spins in our present analysis can move, changing their spatial positions. Such a mobility of the spin-1/2 carrying orphan spins would eventually destabilize the SG order. 

 Another interesting consequence of our observation that orphan spins are mobile, though only in a diffusive manner, could affect the transport properties. Since orphan spins carry both spin and energy, they could give rise to nontrivial spin and thermal transports. In the 1D random-singlet state, the DC spin conductivity was predicted to diverge in the $T\rightarrow 0$ limit \cite{Damle,Damle2}. In the 2D random-singlet state, the diffusive motion of orphan spins could give rise to nontrivial magnetic contributions to the spin and heat transports, {\it e.g.\/}, of the variable-range-hopping type proportional to $\exp[-(T_0/T)^{\frac{1}{D+1}}]=\exp[-(T_0/T)^{\frac{1}{3}}]$ \cite{Mott}. Meanwhile, in the absence of the standard itinerant-type spinons in the 2D random-singlet state, the thermal conduction is expected to be dominated by the phonon contribution. If one assumes that the phonon is scattered by the orphan spin, and the phenomenological argument of the two-level system of Ref.\cite{AndHalVar} is applicable as in the case of the specific heat, the low-$T$ thermal conductivity is proportional to $T^2$, which seems fully consistent with the recent experimental measurements on the organic salt EtMe$_3$Sb[Pd(dmit)$_2$]$_2$ reporting $\sim T^2$ \cite{SYLi} and $\sim T^{1.7}$ \cite{Teillefer} behavior (see also Ref.\cite{Yamashita-reply}).

 In the present paper, we have investigated the properties of the random-singlet state for the case of stronger randomness of $\Delta=1$. This is simply because our numerical method is practically limited to very small systems. Since the random-singlet features, which are nothing but the randomness-induced QSL features, are expected to be most eminent for stronger randomness, being clearly visible even for smaller systems, the stronger random case is best suited to our present numerical study. However, it is also quite possible that the random-singlet-like state is extended to somewhat weaker randomness, but manifest itself there only on longer length scales (larger systems) and at lower temperatures (lower energies) not directly accessible by the present numerical method. On decreasing the randomness $\Delta$ beyond a certain critical value $\Delta_c$, the random-singlet state generally exhibits a quantum phase transition into another phase, {\it e.g.\/}, the two-sublattice \cite{Uematsu2} and the three-sublattice AF phase \cite{Watanabe}, the stripe-type AF phase \cite{Uematsu2}, the VBC-type nonmagnetic state \cite{Uematsu}, or the randomness-irrelevant QSL state \cite{Kawamura}. As mentioned above, in the random-singlet state in the weaker randomness regime of $\Delta\sim \Delta_c$, the random-singlet features appear only on longer length scales (larger systems) and at lower temperatures, while the features of another phase might dominate the shorter-length and higher-temperature behavior. Even in such weaker randomness case, the random-singlet features manifesting themselves on longer length and time scales are adiabatic continuations of those for the stronger randomness case so long as the system does not pass any phase transition point.
% Such random-singlet behaviors observed even for somewhat weaker randomness only at longer length scales and at lower temperatures might be important in certain experimental situations.

\medskip
\noindent {\it Brief survey of the experimental status}

 Now, we wish to discuss the present experimental status from the viewpoint of the randomness-induced QSL state. Some were already given in Refs.\cite{Watanabe,Kawamura,Shimokawa,Uematsu,Uematsu2}. We do not intend to be exhaustive here, but wish to pick up some interesting examples related to the randomness-induced QSL state from both organic and inorganic materials.

 Let us begin with organic materials. As were referred to in \S 1, well-studied examples of organic QSL are $\kappa$-(ET)$_2$Cu$_2$(CN)$_3$ \cite{ET-Kanoda, ET-Kawamoto,ET-Kurosaki,ET-Shimizu,ET-Ohira,ET-Nakazawa,ET-Matsuda,ET-Jawad,ET-Pratt,ET-Goto,ET-Sasaki}, EtMe$_3$Sb[Pd(dmit)$_2$]$_2$ \cite{dmit-Itou,dmit-Tamura,dmit-Itou2,dmit-Matsuda,dmit-Nakazawa,dmit-Itou3,dmit-Watanabe,dmit-Jawad}. QSL features observed in these salts have widely been interpreted as an attribute of clean and homogenous system, {\it e.g.\/}, the one borne by fermionic ``spinons'' with ``spinon Fermi surface''. In contrast, Ref.\cite{Watanabe} pointed out that, since the spin-1/2 in these organic salts spreads over a molecular dimer whose size is an order of magnitude bigger than that in standard inorganic magnets, it could have rich internal degrees of freedom, {\it e.g.\/}, the electric-polarization degrees of freedom, associated with the possible biased charge distribution within a dimer molecule. As both $\kappa$-(ET)$_2$Cu$_2$(CN)$_3$ and EtMe$_3$Sb[Pd(dmit)$_2$]$_2$ are located close to the ferroelectric charge order where the charge-ordered cluster formation and the associated critical slowing-down are to be expected, the QSL state might be the dipole-glass-like state in its charge sector. Indeed, the ac dielectric constant measurements on both $\kappa$-ET \cite{ET-Jawad} and dmit \cite{dmit-Jawad} salts revealed the occurrence of the relaxor-like slowing-down in the QSL regime, suggestive of the random freezing of the dielectric degrees of freedom. Such an inhomogeneous charge (spin) distribution among molecular dimers would give rise to the spatially-modulated random exchange coupling $\{ J_{ij} \}$ between the spin-1/2 on each molecular dimer \cite{Watanabe}.

 In fact, reports of such an inhomogeneous charge and spin distribution can be found in the literature from an earlier stage. Kawamoto {\it et al\/} reported the signature of the inhomogeneous charge distribution in the QSL state of $\kappa$-ET from the $^{13}$C NMR and the in-plane conductivity measurements \cite{ET-Kawamoto}. Shimizu {\it et al\/} reported from the NMR measurements on $\kappa$-ET that the application of magnetic fields in the QSL state induced a spatially inhomogeneous weak magnetic moments \cite{ET-Shimizu}. This observation is fully consistent with our present random-singlet picture, since the application of fields tends to break weaker singlet-dimers, which are located randomly in space, and induce magnetic moments there.

 Magnetic fields generally tend to weaken the strong singlet, break the weak singlet, and polarize the triplet or nearly-free spins, shifting the energy scale of singlets. Reflecting the hierarchical nature of the random-singlet state consisting of the spectrum of singlet-dimers without characteristic energy scale, from strong singlets to weak singlets and eventually to orphan spins, the application of modest magnetic fields would keep at low energies such features of the zero-field spectrum of singlets, except for producing certain amount of spatially-random magnetic moments. So, it would appear as if applied fields erased the high-energy part of singlet-dimers and instead produced the spatially-random magnetic moments.

 Shimizu {\it et al\/} further reported that the nuclear magnetization relaxation obeyed the stretched-exponential form $\sim \exp[-(T_1/T)^\beta]$ at low $T$, indicative of the inhomogeneous spin distribution \cite{ET-Shimizu}. Nakajima {\it et al\/} suggested from the $\mu$SR measurements that the QSL state of $\kappa$-ET is a ``microscopically phase separated'' state \cite{ET-Goto}. The proposed state is inhomogeneous on the microscopic length scale, probably being not far from the inhomogeneous state we are discussing here.

 For the QSL organic salt EtMe$_3$Sb[Pd(dmit)$_2$]$_2$, the NMR measurements by Itou {\it et al\/} also reported the stretched-exponential-type decay of the nuclear magnetization similar to that observed in $\kappa$-ET, indicative of the inhomogeneous spin distribution \cite{dmit-Itou2,dmit-Itou3}. More recently, Yamamoto {\it et al\/} revealed by the IR and Raman spectra measurements that this molecular system shows dynamical charge and lattice fluctuations among molecular dimer, tetramer and octamer configurations due to the competition among these distinct charge-ordered states, leading to the inhomogeneous charge distribution \cite{Yamamoto}. Since the oligomers involved in these fluctuations are big in size, one may expect that the timescale associated with these oligomer transformations might involve slow components, slower than the typical timescale of spin fluctuations in the QSL state. If so, such slowed-down inhomogeneous charge and lattice distributions would provide the spin degrees of freedom the spatially modulated random exchange coupling $\{ J_{ij} \}$, exactly the same situation assumed in the modelling in the present paper and in Refs.\cite{Watanabe,Shimokawa}. Furthermore, for sister dmit salts $X$[Pd(dmit)$_2$]$_2$ ($X=$Me$_4$P, Me$_4$Sb) exhibiting the AF order, Fujiyama and Kato observed by the $^{13}$C NMR measurements the biased distribution of the spin density occurring even in a Pd(dmit)$_2$ monomer molecule between the two ligands on both sides of Pd \cite{Fujiyama}.

 The randomness-induced QSL scenario naturally expects that certain non-QSL magnets, say, the standard AF, could exhibit QSL behavior when randomized artificially. In inorganic materials, certain mixed-crystal magnets, {\it e.g.\/}, Sr$_2$Cu(Te$_{1-x}$W$_{x}$)O$_6$ \cite{Mustonen,Mustonen2,Tanaka} referred to in \S I, is such an example where both end members Sr$_2$CuTeO$_6$ and Sr$_2$CuWO$_6$ exhibit the standard AF order while the mixed crystal Sr$_2$Cu(Te$_{1-x}$W$_{x}$)O$_6$ exhibit a gapless QSL behavior \cite{Mustonen,Mustonen2,Tanaka}.

 In organic salts, a similar case was reported by Furukawa {\it et al\/} for $\kappa$-(ET)$_2$Cu[H(CN)$_2$]Cl, which exhibits the standard AF order in the original state, but upon X-ray irradiation exhibits a gapless QSL behavior with the $T$-linear low-$T$ specific heat \cite{Furukawa}. Since the X-ray irradiation is expected to introduce quenched randomness into the sample, transformation of the AF state to the gapless QSL state upon X-ray irradiation is fully consistent with the randomness-induced QSL scenario.

 In organic salts $\kappa$-(ET)$_2$Cu$_2$(CN)$_3$ and EtMe$_3$Sb[Pd(dmit)$_2$]$_2$, certain amount of extrinsic (quenched) randomness actually exists, {\it i.e.\/}, the random orientation of either CN or NC between two Cu ions in case of $\kappa$-ET salt, and the random orientation of the ethyl group on Sb in case of dmit salt. Such quenched randomness in itself might be too weak to induce the QSL state, but might serve as a ``seed'' in the formation of self-generated charge inhomogeneity mentioned above.

 As a sister compound of well-studied $\kappa$-(ET)$_2$Cu$_2$(CN)$_3$, $\kappa$-(ET)$_2$Ag$_2$(CN)$_3$ was recently synthesized by Shimizu {\it et al\/}, which was also found to exhibit the gapless QSL behavior \cite{Shimizu}. 

 In addition to $\kappa$-(ET)$_2$Cu$_2$(CN)$_3$ and EtMe$_3$Sb[Pd(dmit)$_2$]$_2$ consisting of alternating anion-cation layers, a different type of gapless QSL material $\kappa$-H$_3$(Cat-EDT-TTF)$_2$ was reported \cite{Isono1,Isono2,Ueda}. While the spin-1/2 spreads over a big Cat molecule forming the triangular lattice as in $\kappa$-(ET)$_2$Cu$_2$(CN)$_3$ and EtMe$_3$Sb[Pd(dmit)$_2$]$_2$, neighboring Cat molecules are coupled via the hydrogen bond, a unique feature of $\kappa$-H$_3$(Cat-EDT-TTF)$_2$. Theoretical calculation suggests that the effective potential felt by the proton is nearly flat along the hydrogen bond without clear double minima usually formed at the hydrogen bond \cite{Tsumuraya}, so that the proton would show large displacement along the hydrogen bond in its QSL state. While in the picture of Refs.\cite{Isono1,Isono2,Ueda} its QSL state was borne by simultaneous dynamical fluctuations of proton positions and of $\pi$-electron spins, since the proton is much heavier than the electron, it seems more likely to the present authors that the proton movement slows down on decreasing the temperature, and nearly stop at the time scale of $\pi$-electron-spin fluctuations at spatially random positions along the hydrogen bonds due to the almost flat character of its potential (adiabatic approximation). Then, these almost stopped protons attract the spin-1/2 carrying $\pi$-electrons on the Cat molecules on both sides of the hydrogen bond, spatially modifying the effective magnetic coupling between the spin-1/2 on the neighboring Cat molecules at the time scale of $\pi$-electron-spin fluctuations, exactly the situation leading to the random-singlet state. Thus, $\kappa$-H$_3$(Cat-EDT-TTF)$_2$ is likely to be an another example of the gapless QSL induced by the self-generated intrinsic randomness or inhomogeneity.
 
 At this point, a comment concerning the nature of the {\it intrinsic randomness\/} might be worthwhile. Generally speaking, self-generated intrinsic randomness is expected to arise when certain other degrees of freedom in solids, {\it e.g.\/}, the charge degrees of freedom, provide a spatially random effective interaction to the spin degrees of freedom. The quenched (or adiabatic) approximation assumed in the present approach would be a better approximation when the separation of times scales develops between the spin and the other degrees of freedom, {\it i.e.\/}, the time scale of the other degrees of freedom $\tau_{other}$ gets much longer than that of spin fluctuations $\tau_{spin}$, $\tau_{other}\gg \tau_{spin}$. In the opposite limit of $\tau_{other}\ll \tau_{spin}$, fluctuations of the other degrees of freedom will be averaged out, and their mean (or most probable) values would suffice in considering the spin state.

 Meanwhile, the extent of the time-scale separation is a case-dependent relative issue. When the two time scales are not much different $\tau_{spin}\sim \tau_{other}$, the spin and the other degrees of freedom would fluctuate simultaneously in a cooperative manner, forming the complex. This is also a very interesting situation. In such a situation, one needs to tackle both the spin and the other degrees of freedom simultaneously on equal footing, and our present modelling would become poorer. Even in that case, however, the adiabatic approach assuming $\tau_{spin}\ll \tau_{other}$ could still provide some useful information about what happens in the limit opposite to the averaged-out case of $\tau_{spin}\gg \tau_{other}$, especially under the condition that the full theoretical analysis properly describing the $\tau_{spin}\sim \tau_{other}$ situation is much tougher.

 As an example of the organic QSL induced by {\it extrinsic\/} randomness, organic radical compound Zn(hfac)$_2$(A$_x$B$_{1-x}$) forming an $s=1/2$  honeycomb lattice was  reported by Yamaguchi {\it et al\/} \cite{Yamaguchi}. In this compound, the quenched randomness of the intermolecular exchange interactions arises from the two different regioisomers, A and B, that randomly align in the crystal. The magnetic and thermodynamic experimental results indicate the gapless QSL behavior with the $T$-linear specific heat and the gapless low-$T$ susceptibility with a Curie-like tail. Similar organic radical systems can easily introduce an extrinsic randomness into the spin lattices, so that the randomness-induced QSL could be realized on various lattices.

 As such, we now have considerable amount of experimental evidence of the randomness (inhomogeneity)-induced QSL in organic salts. The underlying randomness is either of intrinsic origin self-generated in clean material via the coupling to other degrees of freedom, or of extrinsic origin, {\it i.e.\/}, the quenched randomness.

 Of course, target materials are not necessarily limited to organic materials, but also include a variety of inorganic materials as introduced in \S I. Those include a kagome antiferromagnet herbertsmithite ZnCu$_3$(OH)$_6$Cl$_2$ and a pyrochlore antiferromagnet Lu$_2$Mo$_2$O$_5$N$_2$ with geometrical frustration, as well as a square-lattice magnet Sr$_2$Cu(Te$_{1-x}$W$_{x}$)O$_6$ and a honeycomb-lattice magnet 6HB-Ba$_3$NiSb$_2$O$_9$ where the frustration is borne by the competition of the NN and the NNN interactions.

 In addition to these examples, it was reported that the $s=1/2$ triangular-lattice antiferromagnet Cs$_2$Cu(Br$_{1-x}$Cl$_x$)$_4$, which exhibits the standard AF order in the vicinity of $x=0$ and $x=1$, exhibits a QSL behavior for an intermediate $x$-range \cite{Ono}. Indeed, since considerable amount of randomness associated with the random mixing of Cl$^-$ and Br$^-$ is expected in this material, a randomness-induced QSL state is certainly expected for an intermediate $x$, consistently with the experiment. A kagome antiferromagnet Zn-brochantite, ZnCu$_3$(OH)$_6$SO$_4$ containing a significant amount of randomness, was reported to exhibit a gapless QSL behavior with the $T$-linear low-$T$ specific heat \cite{Li,Gomilsek,Gomilsek2}, and could also be an example of the randomness-induced gapless QSL state.

 Another interesting candidate material might be LiZn$_2$Mo$_3$O$_8$, which was reported to exhibit a gapless QSL behavior \cite{Sheckelton,Sheckelton2,Mourigal}. This material consists of a triangular network of Mo$_3$O$_{18}$ cluster units, and $s=1/2$ exists on each molecular unit spreading over three Mo atoms. In fact, this is an analogous situation to the one in organic salts discussed above, where $s=1/2$ exists spreading over a big molecular dimer forming the triangular lattice. Thus, one may speculate that, on decreasing the temperature, the motion of the $s=1/2$-carrying electron slows down in a Mo$_3$O$_{18}$ cluster, leading to the bias in its location among three Mo atoms in a spatially random manner, from one molecular unit to the other. Then, a slowing-down of spatially random local electric polarization would arise without any macroscopic polarization, which would be detectable via the relaxor-like response in the ac dielectric constant, as was observed in organic salts \cite{ET-Jawad,dmit-Jawad}. To the authors' knowledge, there is no experimental evidence of such behavior at the present stage, but it may be interesting to test such a possibility as a possible route to the observed gapless QSL behavior in LiZn$_2$Mo$_3$O$_8$.

 Similar situation is also expected in another inorganic QSL candidate, 1T-TaS$_2$ \cite{Kratochvilova,Yu,Arcon,Ribak,Roy}. At low temperatures, this layered material exhibits a commensurate CDW order forming a periodic array of star-of-Dvaid arrangements with 13 Ta atoms as a cluster unit, and the emergent spin $1/2$ exists at this cluster unit spreading over the 13 Ta atoms in the cluster, with large in-plane displacements of 12 surrounding Ta atoms toward the central Ta atom in the star-of-David cluster. This is again an analogous situation to the ones in organic salts and in LiZn$_2$Mo$_3$O$_8$ discussed above. The gapless QSL behavior characterized by the $T$-linear low-$T$ specific heat and the almost constant susceptibility followed by an intrinsic Curie-like tail at lower temperatures was observed experimentally \cite{Kratochvilova,Ribak}. Interestingly, the NQR measurements showed that the NQR spectrum in the QSL state was fully split, indicating the local deformations of the 12 surrounding Ta atoms \cite{Arcon}. The $1/T_1$ measurements revealed the growth of the local-field inhomogeneity, suggesting the occurrence of a highly inhomogeneous magnetic state \cite{Arcon}. All these experimental observations suggest that this inhomogeneous QSL-like state is most likely the 2D frustrated random-singlet state observed for the random $s=1/2$ AF Heisenberg model on the triangular lattice \cite{Watanabe,Shimokawa}.

 Concerning the origin of the randomness, while the stacking disorder or the slight off-stoichiometry was suggested  \cite{Arcon}, we wish to propose here a scenario similar to the one proposed for organic salts \cite{Watanabe}, {\it i.e.\/}, the motion of the $s=1/2$-carrying electron slows down in each star-of-David cluster, leading to the bias in its position from the center in a spatially inhomogeneous manner, from one cluster to the other. Such a spatial charge inhomogeneity would give rise to the spatially random modulation of the exchange coupling $J_{ij}$, leading to the 2D random singlet state due to the self-generated intrinsic randomness, as discussed in the present paper and in Refs.\cite{Watanabe}. It would also be interesting to perform the ac dielectric constant measurement in search for the relaxor-like response due to the possible inhomogeneous slowing-down of the $s=1/2$-carrying charge or dielectric degrees of freedom.

 Recently, a rare-earth triangular-lattice antiferromagnet YbMgGaO$_4$ where Yb$^{3+}$ bears an effective $s=1/2$ was reported to exhibit a gapless QSL-like behavior \cite{Yb-Li,Yb-Li2,Yb-Xu,Yb-Li3,Yb-Shen,Yb-Li4,Yb-Paddison,Yb-Zhang,Yb-Toth,Yb-Li5}. The existence of the quenched randomness due to the inter-layer nonmagnetic Ga$^{3+}$/Mg$^{2+}$ mixing was also indicated experimentally \cite{Yb-Li,Yb-Li4}. In contrast to the QSL magnets discussed in the present paper, which are more or less isotropic with relatively weak magnetic anisotropy, the exchange interaction in this material is highly anisotropic with significant off-diagonal components due to the strong spin-orbit coupling \cite{Yb-Li,Yb-Li2,Yb-Toth,Yb-Paddison,Yb-Zhang}. Indeed, while all QSL magnets discussed in the present paper exhibit the $T$-linear low-$T$ specific heat without a sign of the SG freezing, YbMgGaO$_4$ was reported to exhibit the low-$T$ specific heat proportional to $T^\alpha$ with a fractional power $0.6\lesssim \alpha \lesssim 0.74$ \cite{Yb-Li,Yb-Xu}, and a sign of the spin-glass-like freezing observed in the ac susceptibility for some samples \cite{Yb-Ma}. Yet, in other measurements, no sign of the spin freezing was observed down to low temperatures \cite{Yb-Li3,Yb-Shen,Yb-Paddison,Yb-Li5}. Thus, the nature of the QSL of YbMgGaO$_4$ remains controversial. 
 Although an analogy to the random-singlet state of 2D frustrated magnets discussed in Refs.\cite{Watanabe,Kawamura,Shimokawa,Uematsu,Uematsu2} was not noticed in the literature on YbMgGaO$_4$ \cite{Yb-Li2,Yb-Xu,Yb-Li3,Yb-Shen,Yb-Li4,Yb-Paddison,Yb-Zhang,Yb-Toth,Yb-Li5,Yb-Ma}, certain essential features, {\it e.g.\/}, the occurrence of the gapless QSL on the frustrated triangular lattice \cite{Watanabe} with the continuous gapless spectrum of the dynamical spin structure \cite{Shimokawa}, as well as the existence and the potential importance of the randomness, are in common. At the same time, an apparent difference also exists, {\it e.g.\/} the low-$T$ specific heat exhibits the $T^\alpha$ ($0.6\lesssim \alpha \lesssim 0.74$) behavior in YbMgGaO$_4$ in contrast to the $T$-linear behavior in the 2D random-singlet state of Refs.\cite{Watanabe,Kawamura,Shimokawa,Uematsu,Uematsu2}. Thus, to clarify the true relationship between the QSL-like state observed in anisotropic YbMgGaO$_4$ and the 2D frustrated random-singlet state discussed for Heisenberg-like isotropic magnets would be informative.

In this way, randomness-induced gapless QSL behavior prevail both in inorganic and organic materials. Although the gapless QSL now observed in many organic and inorganic magnets has been discussed mostly in the context of ``spinons'' or ``spinon Fermi surface'' essentially as an attribute of clean and homogenous system, the present authors believe that, for many of them, reconsideration of their QSL properties as the randomness or the inhomogeneity induced ones would be necessary and worthwhile. They appear to form a novel class of randomness-induced magnetic state, where both strong quantum fluctuations and frustration play a crucial role. Indeed, cooperation of strong quantum fluctuations, frustration and randomness, sometimes with the help of the coupling between the spin and other degrees of freedom in solids, opens a route to a new state of matter. Finally, let us emphasize again that the randomness or inhomogeneity, a fact of life, is not just an unnecessary contamination nor a dirty extra, but often provides us with rich and interesting new phenomena. The randomness-induced QSL is most probably such an example.

\begin{acknowledgements}
The authors are thankful to Dr. K. Aoyama, Dr. T. Shimokawa, Dr. T. Hikihara, Dr. T. Okubo, Dr. A.W. Sandvik, Dr. H. Yamaguchi, Dr. Y. Shimizu, Dr. R. Kato  and Dr. H. Fukuyama for useful discussion and comments. This study was supported by JSPS KAKENHI Grants No. 17H06137. Our code was based on TITPACK Ver.2 coded by H. Nishimori. We are thankful to ISSP, the University of Tokyo, and to YITP, Kyoto University, for providing us with CPU time.
\end{acknowledgements}

\end{document}